\documentclass[11pt,preprint]{aastex}

\input epsf.tex
\newcommand{\mincir}{\raise -2.truept\hbox{\rlap{\hbox{$\sim$}}\raise5.truept
\hbox{$?$}\ }}
\newcommand{\gr}{\kern 2pt\hbox{}^\circ{\kern -2pt K}} %  ====? GRADI KELVIN
\newcommand{\magcir}{\raise -2.truept\hbox{\rlap{\hbox{$\sim$}}\raise5.truept
\hbox{$?$}\ }}

\newcommand{\Om}{\Omega}
\newcommand{\om}{\omega}
\newcommand{\si}{\sigma}
\newcommand{\be}{\begin{equation}}
\newcommand{\ee}{\end{equation}}
\newcommand{\bea}{\begin{eqnarray}}
\newcommand{\eea}{\end{eqnarray}}
\newcommand{\De}{\Delta}

\newcommand{\bn}{{\bf n}}
\newcommand{\bx}{{\bf x}}
\newcommand{\bk}{{\bf k}}
\newcommand{\etal}{{et al.}}

\newcommand{\eg}{{\it e.g.}}

\shorttitle{Acoustic peaks and dips in the CMB power spectrum: 
	theory and observations}
\shortauthors{Durrer et al.}

\begin{document}

\title{Acoustic peaks and dips in the CMB power spectrum:\\
observational data and cosmological constraints}

\author{R.~Durrer\altaffilmark{1}}
\affil{School of Natural Sciences, Institute for 
	Advanced Study, Einstein Drive, Princeton, NJ 08540}
  \author{B.~Novosyadlyj \altaffilmark{2} and S. Apunevych}
\affil{Astronomical Observatory of Ivan Franko L'viv  National University, Kyryla and
Mephodia str.8, 79005, L'viv, Ukraine}
\altaffiltext{1}{Department de Physique Th\'eorique, Universit\'e de Gen\`eve,
Quai Ernest Ansermet 24, CH-1211 Gen\`eve 4, Switzerland}
\altaffiltext{2}{on leave from Department de Physique Th\'eorique, 
Universit\'e de Gen\`eve, Switzerland}

\begin{abstract}
The locations and amplitudes of three acoustic peaks and two dips 
in  the Boomerang, MAXIMA and DASI measurements of 
the cosmic microwave background (CMB) anisotropy power spectra
as well as their statistical confidence levels are determined  in a 
model-independent way. It is shown that the Boomerang-2001 
data \cite{boom01} fixes the location and amplitude of the first 
acoustic peak at more than $3\sigma$ confidence. The next 
two peaks and dips are determined at a confidence level above
 $1\sigma$ but below $2\sigma$.
The locations and amplitudes of the first three peaks and two dips are 
$\ell_{p_1}=212\pm 17$, $A_{p_1}=5426\pm 1218\;\mu K^2$, 
$\ell_{d_1}=413\pm 50$, $A_{d_1}=1960\pm  503\;\mu K^2$, 
$\ell_{p_2}=544\pm 56$, $A_{p_2}=2266\pm  607\;\mu K^2$,
$\ell_{d_2}=746\pm 89$, $A_{d_2}=1605\pm  650\;\mu K^2$,
$\ell_{p_3}=843\pm 35$, $A_{p_3}=2077\pm  876\;\mu K^2$ 
respectively ($1\sigma$ errors include statistical and systematic errors).
 The MAXIMA and DASI experiments give similar values
for the extrema which they determine. For MAXIMA these are the  1st and 
3rd peaks, for DASI the 1st and 2nd peaks and the 1st dip. Moreover, 
the locations and amplitudes of the extrema determined from the combined
data of all experiments are quite close to the
corresponding values extracted from the Boomerang data alone.

In order to use  these data in a fast search for cosmological parameters 
an accurate analytic approximation to calculate CMB peak and dip positions
and amplitudes in mixed dark matter models with cosmological constant
and curvature is derived and tested. 
 
The determined cosmological parameters from the CMB acoustic
extrema data show good agreement with other determinations, especially
with the baryon content as deduced from standard nucleosynthesis 
constraints \cite{burles}.
These data supplemented by constraints from direct measurements of 
some cosmological parameters and data on large scale structure (LSS) 
lead to a best-fit model which agrees with practically all the used 
experimental data within $1\sigma$. The best-fit parameters are:
$\Omega_{\Lambda}=0.64^{+0.14}_{-0.27}$, $\Omega_{m}=0.36^{+0.21}_{-0.11}$,
$\Omega_b=0.047^{+0.093}_{-0.024}$, $n_s=1.0^{+0.59}_{-0.17}$, $h=0.65^{+0.35}_{-0.27}$
and $\tau_c=0.15^{+0.95}_{-0.15}$. The best-fit values of $\Omega_{\nu}$ and $T/S$
are close to zero, their 1$\sigma$ upper limits are  $0.17$ and $1.7$ respectively. 

\end{abstract}


\section{Introduction}
The new data on the cosmic microwave background (CMB)  temperature 
anisotropy obtained in the Boomerang (de Bernardis et al. 2000; Netterfield
et al. 2001), Maxima~I 
\cite{maxima00,maxima01} and DASI \cite{dasi} experiments 
provide relatively accurate measurements of the CMB anisotropies up to 
$\ell \sim 1000$.  Boomerang is a long duration balloon (LDB) flight 
around the South Pole, MAXIMA is a balloon flight from  Palestine, 
Texas, and DASI is an interferometer experiment. The mutual 
agreement of such divers experiments within statistical uncertainties 
is very reassuring.

After a correction of the first results from Boomerang \cite{boom00} by 
Netterfield et al. (2001),  these 
measurements are in astounding agreement with the simplest
flat adiabatic purely scalar model of structure formation. The best fit 
cosmological parameters obtained coincide with other, completely 
independent determinations, like \eg~ the baryon density parameter 
predicted by nucleosynthesis \cite{burles}. 

CMB anisotropies can be calculated within linear perturbation theory in a 
multi-component universe. These calculations are very well established 
and allow accurate predictions of the CMB power spectrum for a given 
model of initial perturbations and given cosmological parameters. All 
the calculations are linear and very well controlled. Publicly available 
codes, \eg~ CMBfast \cite{cmbfast} provide 1\%  accurate results for a 
given model within two minutes of CPU time on an ordinary PC. Due to 
these advantages,  CMB temperature fluctuation data are extremely 
valuable for testing theoretical models of structure formation and for the
determination of cosmological parameters.

Nevertheless, the efficiency of  parameter determination using codes like 
CMBfast to compute the temperature anisotropy spectrum for each model
has several problems: 
1) The complete set of observational data of the current state and the 
early history of the Universe is described by models with at least six 
parameters. The implementation of CMBfast-like codes into search 
procedures for best fits in high dimensional parameter spaces consumes 
too much CPU time even for the most advanced computers due to the 
necessity to carry out numerical integration of the  Einstein-Boltzmann 
system of equations which describe the evolution of temperature and 
density perturbations of each component through the decoupling epoch. 
2)  The CMB power spectrum alone has several more or less exact 
degeneracies in parameter space (see \eg ~Efstathiou \& ~Bond (1999))  
which can only 
be reduced substantially or removed completely if other data sets, 
\eg~ galaxy clustering data, corresponding to different scales and 
redshifts, are combined with CMB measurements. The results and 
especially the error bars which are obtained from search procedures 
using different classes of cosmological observations with different 
quality and different statistical properties are difficult to 
interpret.  

Several groups have overcome the first problem by computing a grid of CMB 
anisotropy spectra in the space of models and interpolating between them
to obtain the spectra for intermediate values of the parameters (see 
Tegmark et al. (2001); Lange et al. (2001); Balbi et al. (2000); de 
Bernardis et al. (2001); Wang et al. (2001) and references therein). 

Here we propose an alternative method:
The  CMB angular power spectrum obtained by COBE \cite{cobe92}, 
Boomerang, MAXIMA-1 and DASI  has well defined statistical and systematic 
errors in the range of scales from quadrupole up to the spherical harmonic
$\ell\sim 1000$  and the present data can be represented by a few dozen
uncorrelated measurements. Practically the same information
is contained in a few characteristics such as the amplitude and 
inclination of the power spectrum at COBE scale and the amplitudes and 
locations of the observed acoustic peaks and dips. 
Indeed, it was shown (see \cite{hs95,hw96,efs99,dl01} and references therein)
that the sensitivity of the locations and amplitudes of the extrema, 
especially of the peaks, to cosmological parameters is the same as their
sensitivity to the flat band powers, the $C_l$'s, from presently available
data.  This is not surprising: the extrema have an obvious physical 
interpretation and have the largest weight among data points on 
CMB power spectrum as a result of a minimal ratio of error to value.
Hu et al. (2001) have shown that most of the information from the Boomerang 
and MAXIMA-1 data sets can be compressed into four observables: amplitude 
and location of 1-st peak, and amplitudes of 2-nd and 3-rd peak.

The first three 
acoustic peaks and the two dips indicated by the above mentioned 
experiments and the COBE large scale data can be presented by not more 
than 12 experimental points. If we use the approach by 
Bunn \& White (1997) for the four year COBE data  and the data on 
acoustic peaks, we have 7 experimental values to compare with 
theoretical models. Each of them  can be calculated by analytical or 
semi-analytical methods. This enables us to study present CMB
data in a very fast search procedure for multicomponent models.

Even though, in principle, a measurement of the entire $C_\ell$ spectrum
contains of course more informations than the position and amplitudes
of its peaks and dips, with present errorbars, this additional informations
seems to be quite modest.

The goal of this paper is to use these main characteristics of the CMB 
power spectrum to determine cosmological parameters. To do this we have to
accomplish the following steps: 1) to locate the positions and 
amplitudes of three peaks and two dips as well as determining their 
error bars from experimental data, 
2) to derive accurate analytical approximations to calculate these 
positions and amplitudes and test them by full numerical 
calculations. We also derive an accurate and fast semi-analytical method 
to normalize the  power spectrum to the 4-year COBE data.
Such analytical approximations have been derived in the past for the 
matter power spectrum \cite{eh98,eh99,ndl99} and for the Sachs-Wolfe 
part of the CMB anisotropy spectrum (Kofman \& Starobinsky 1985; Apunevych 
\& Novosyadlyj 2000). Here we derive 
an analogous approximation for the acoustic part of the CMB anisotropy 
spectrum by improving an approximation proposed by Efstathiou \& ~Bond
(1999).

The outline of the paper is as follows. In Section 2 we determine the 
locations and amplitudes of the 1st, 2nd and 3rd acoustic peak as well as 
1st and 2nd dip and their confidence levels using 
the published data on the CMB angular power spectrum from 
Boomerang \cite{boom01}, MAXIMA-1 \cite{maxima01} and DASI (Halverson et 
al., 2001).
Analytical approximations for the positions and amplitudes of  the 
acoustic peaks and dips are described in Section 3. A new method for an 
accurate and fast COBE normalization is also presented in this section.
Details are given in two Appendices. 
Our search procedure to determine cosmological parameters along with
the discussion of the results is presented in Section 4. In Section 5.
we draw conclusions.

\section{Peaks and dips in the CMB power spectrum: experimental data}

We have to determine the locations and amplitudes of acoustic peaks and 
dips as well as their uncertainties in the data of the
angular power spectrum of CMB temperature fluctuations obtained in the 
Boomerang \cite{boom01}, MAXIMA-1 \cite{maxima01} and  DASI (Halverson et 
al. 2001). 
We carry out a model-independent analysis of the peaks and dips in the power
spectra for each experiment separately, as well as using all data points 
jointly.   
  
 \subsection{Boomerang-2001}

A model-independent determination of peak and dip locations and amplitudes in 
the Boomerang data \cite{boom01} has been carried out recently by de 
Bernardis \etal ~(2001).  Our approach is based on a conceptually somewhat 
different method, especially in the determination of statistical errors.

\begin{figure}
\plotone{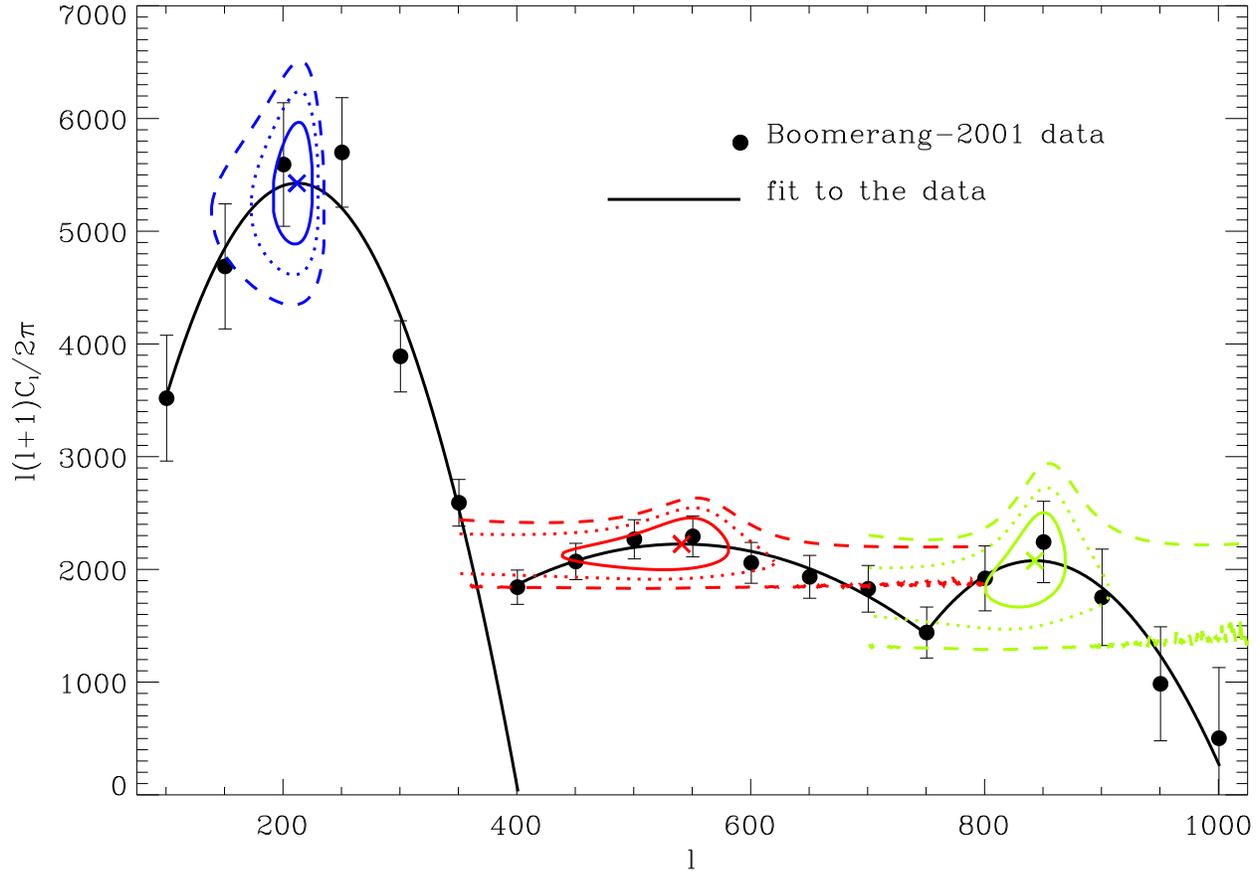}
\caption{Best-fit parabolas for the acoustic peaks of the Boomerang-2001 CMB 
power spectrum ($\chi_{min}^2=2.59$ for the 1st peak, $\chi_{min}^2=0.92$ for 
the 2nd peak and $\chi_{min}^2=0.65$ for the 3rd peak) as well as 
$1$ (solid), $2$ (dotted) and $3\sigma$ (dashed) contours for their 
locations and amplitudes are shown. The crosses indicate the top of the
best-fit parabolas. The contours limit the regions in the  
 $\left(\ell, \ell(\ell+1)C_{\ell}/2\pi\right)$ plane which contain the 
tops of parabolas with 
$\Delta\chi^2=3.53, \;8.02, \;14.2$ for the 1st and 3rd peaks ($N_f=3$) and 
$\Delta\chi^2=5.89, \;11.3, \;18.2$ for the 2nd peak ($N_f=5$).   }
\label{peaks}
\end{figure}

At first, mainly for comparison of our results with  de Bernardis \etal 
~(2001),
we fit the peaks in the Boomerang CMB power spectrum by curves of second 
order (parabolas) as  shown in Fig.~\ref{peaks}. The six experimental points 
($N_{exp}=6$) in the range $100\le \ell \le 350$, which  trace the first 
acoustic peak, are well approximated by a three parameter curve 
($N_{par}=3$). Hence,  the number of degrees of freedom 
$N_f=N_{exp}-N_{par}$ 
for the determination of these parameters is $N_f= 3$. The best-fit
parabola has $\chi_{min}^2=2.6$. Its extremum, located at 
$\ell_{p_1}=212$ and  
$A_{p_1}=5426\;\mu K^2$, is the best-fit location and 
amplitude of the first acoustic peak. We estimate the statistical error 
in the following way. Varying the 3 parameters of the fitting 
curve so that  $\chi^2-\chi_{min}^2\le 3.53$ the maxima of the parabolas 
define the $1\sigma$ range of positions and amplitudes of the first 
acoustic peak in the  plane $\left(\ell, \ell(\ell+1)C_{\ell}/2\pi\right)$. 
The boundary of this region determines the statistical $1\sigma$ errors 
for the location and amplitude of the first acoustic peak. We obtain     
$$\ell_{p_1}=212^{+13}_{-20}, ~~ A_{p_1}=5426^{+540}_{-539}\;\mu K^2.$$
In Fig.~\ref{peaks} the $1\sigma$,  
$2\sigma$ (the boundary of the region with $\chi^2-\chi_{min}^2\le  
8.02$) and $3\sigma$ ($\chi^2-\chi_{min}^2\le  14.2$) contours are shown 
in the plane  $\left(\ell, \ell(\ell+1)C_{\ell}/2\pi\right)$. All contours 
for the first
acoustic peak are closed. This shows that the Boomerang-2001 data  
prove the existence of a first peak at a confidence level higher than 
$3\sigma$. The values $\Delta\chi^2=3.53$, 8.02 and 14.2, given by the  
incomplete Gamma function, $Q(N_f,\Delta\chi^2/2)=1-0.683,~ 1-0.954$ and 
$1-0.9973$, correspond to 68.3\%, 94.5\% and 99.73\% confidence levels 
respectively for a Gaussian likelihood of $N_f=3$ degrees of freedom.
These levels which depend on $N_f$ and thus on the number of independent 
data points (which we just took at face value from Netterfield \etal 
~(2001)) define the regions within which the maxima of parabolas leading 
to the data points lie with a probability $\ge p$, where $p=0.68,~0.954$ 
and $0.9973$ for $1$-, 
$2$- and $3$-$\sigma$ contours respectively. The same method for the
second peak using the eight experimental points in the range 
$400\le \ell \le750$, hence $N_f=5$, gives    
$$ \ell_{p_2}=541^{+40}_{-102},~~  A_{p_2}=2225^{+231}_{-227}\;\mu K^2.$$
For the third peak ($750\le \ell\le 1000$, 6 experimental points, 
$N_f=3$) we obtain
$$ \ell_{p_3}=843^{+25}_{-42},~~  A_{p_3}=2077^{+426}_{-412}\;\mu K^2.$$

For the second and third peaks, the $2$- and $3$-$\sigma$ contours are open 
as shown in Fig.~\ref{peaks}. This means  that the Boomerang experiment 
indicates the 2nd and 3rd acoustic peaks at a confidence level higher than 
$1$- but lower than $2$-$\sigma$. This is in disagreement with the result 
obtained in \cite{ber01}.  Formally the disagreement
consists in the fact that de Bernardis \etal~ (2001) set $N_f=2$ for all 
peaks and dips (see paragraph 3.1.2 of their paper) leading to different 
values of $\Delta\chi^2$ for the $1$-, $2$- and $3$-$\sigma$ contours. They
argue that there are two free parameters, namely the height and the 
position of the peak. The 'philosophy' of the two approaches is somewhat 
different: 
While our contours limit the probability that the given data is measured if 
the correct theoretical curve has the peak position and amplitude inside 
the contour, in their approach the contours limit the probability that the 
given best fit parabola leads to data with peak position and amplitude 
inside the contour. In other words, while they compare a given parabola 
to the best fit curve, we compare it to the data. In that sense we think 
that the closed $2$-$\sigma$ contours of de Bernardis \etal (2001) do not
prove the existence of the secondary peaks at the 2-$\sigma$ level. 
Of course our approach has a problem as well: It relies 
on the data points being independent. If they are not, the number of 
degrees of freedom should be reduced.

The same procedure can be applied for the amplitudes and positions of the 
two dips between the peaks.

\begin{figure}
\plotone{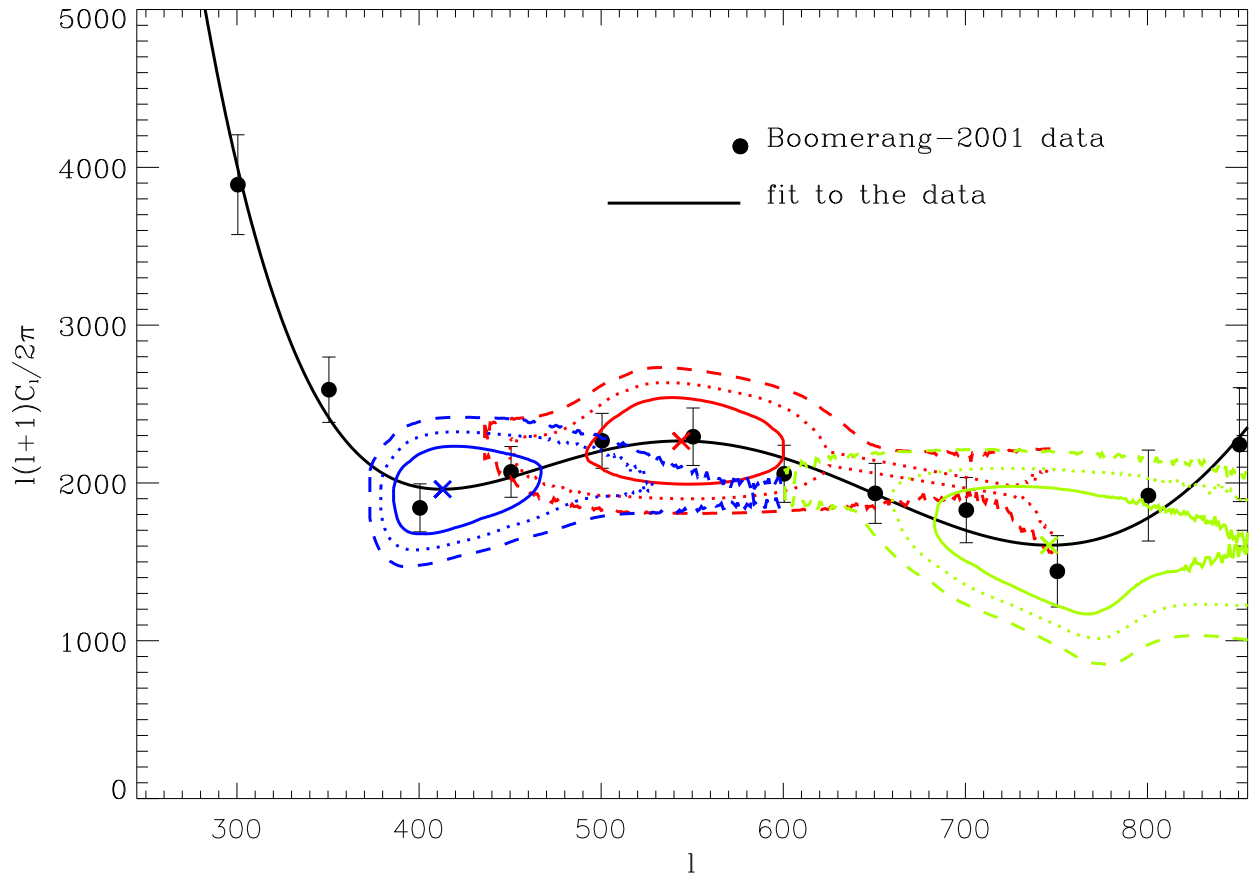}
\caption{The best polynomial fit for the Boomerang-2001 CMB power spectrum 
in the range of the 1st dip, 2nd peak and 2nd dip ($\chi_{min}^2=3.26$), and 
the $1,\;2\;{\rm and}\;3\sigma$ contours for their locations and  amplitudes. 
The crosses indicate the positive (dips) and negative (peak) curvature 
extrema. The contours limit the regions in the  $\left(\ell, 
\ell(\ell+1)C_{\ell}/2\pi\right)$ plane containing the corresponding 
extrema of polynomial fits with 
$\Delta\chi^2=7.04,\;12.8,\;20.1$ ($N_f=6$).}
\label{dips}
\end{figure}

There is also a slight logical problem in
the results presented in Fig.~2 of de Bernardis \etal (2001):
If 2nd and 3rd peaks are established at $2\sigma$ C.L., then the 2nd dip 
should be determined at the same C.L.; but  even the 
$1\sigma$  contour for the position of the second dip is not closed.
Also the position of the first dip is actually fixed by a single data point 
at $\ell=450$ as can be seen from Fig.~3 in \cite{ber01}.
In order to remove these problems we approximate the experimental points 
in the range $250\le \ell\le 850$ (12 experimental points) by one single 
fifth order polynomial (6 free parameters). The number of degrees of 
freedom $N_f=6$. The best-fit curve with $\chi_{min}^2=3.26$  is presented 
in Fig.~\ref{dips}. Its 6 coefficients are 
$a_0=1.13\cdot 10^5$, $a_1=-944.7$, $a_2=3.106$, $a_3=-4.92\cdot 10^{-3}$, 
$a_4=3.75\cdot 10^{-6}$, $a_5=-1.099\cdot 10^{-9}$. This method allows us
to determine the locations and amplitudes of both dips and of the second 
peak by taking into account the relatively prominent raises to the third and
especially to the first peak.  The local extrema of the polynomial best-fit
give the following locations and amplitudes of the 1st and 2nd dip 
(positive curvature extrema) and the 2nd peak (negative 
curvature) between them. The $1\sigma$ error bars are determined as above:
  $$ \ell_{d_1}=413^{+54}_{-27},\;\;\;  A_{d_1}=1960^{+272}_{-282}\;\mu K^2,$$
  $$ \ell_{p_2}=544^{+56}_{-52},\;\;\;  A_{p_2}=2266^{+275}_{-274}\;\mu K^2,$$
  $$ \ell_{d_2}=746^{+114}_{-63},\;\;\;  A_{d_2}=1605^{+373}_{-436}\;\mu K^2.$$
The results discussed here are presented in Table~\ref{tab_boom} and 
shown in Figs.~\ref{peaks} and \ref{dips}.
In Fig.~\ref{dips} also the  $1\sigma$ ($\Delta\chi^2=7.04$), 
$2\sigma$ ($\Delta\chi^2=12.8$) and  $3\sigma$ ($\Delta\chi^2=20.1$) 
confidence contours are shown.
The 1-$\si$ contours for all peaks and dips are now closed.
The $2\sigma$ contour for the 2nd peak
 has a 'corridor' connecting it with the 3rd peak. As we have noted before
(see Fig.~\ref{peaks}), the $2\sigma$ confidence contour for the 3rd 
peak is also open towards low $\ell$. This implies that we can not 
establish the second peak   at $2\sigma$ C.L. The probability of its 
location is spread out over the  entire range $450\le \ell \le 920$. 
Therefore at   $2\sigma$ C.L. we can not state whether the 
Boomerang-2001 results indicate a second peak without a third, or third 
without a second or both. We only can
state at $2\sigma$ C.L. that there are one or two  negative curvature 
extrema of  the function $\ell(\ell+1)C_\ell/2\pi$ situated in the range
$450\le \ell \le 920$ with amplitude in the range 
$1500\le \ell(\ell+1)C_\ell/2\pi \le 2700 \mu K^2$. Furthermore, there is 
one secondary peak present in this $\ell$ range at $2\sigma$ confidence. 
Now, the contours 
for the dips are in logical agreement with the information about the 
second and third peaks.  If at  $2\sigma$ C.L. the 2nd peak can be at 
the range of  the location of 3rd one, then the 1st dip will move to 
$\ell\sim 520$. Its  $2\sigma$ contour is  closed since the 3rd peak  
has a closed  $2\sigma$ C.L. contour at the high $\ell$ side. On the 
contrary, the 2nd dip is open at high $\ell$ as it disappears when the  
'second' peak disappears and the  'third' peak becomes the second. 
The $3\sigma$ contours for the 2nd and 3rd peak as well as for 1st and 
2nd dips are open in the direction of high $\ell$. 

So far we discussed only the statistical errors. The Boomerang LDB 
measurements have two systematic errors: $20\%$ calibration uncertainty 
and beam width uncertainty leading to scale-dependent correlated
uncertainties in the determination of the power spectrum (Netterfield 
et al. 2001). 
The calibration error results in the same relative error for all data 
points and can be taken into account easily. The beam width uncertainty 
which induces an error which becomes larger at higher values of $\ell$,
needs more care.

\begin{figure}[ht]
\plotone{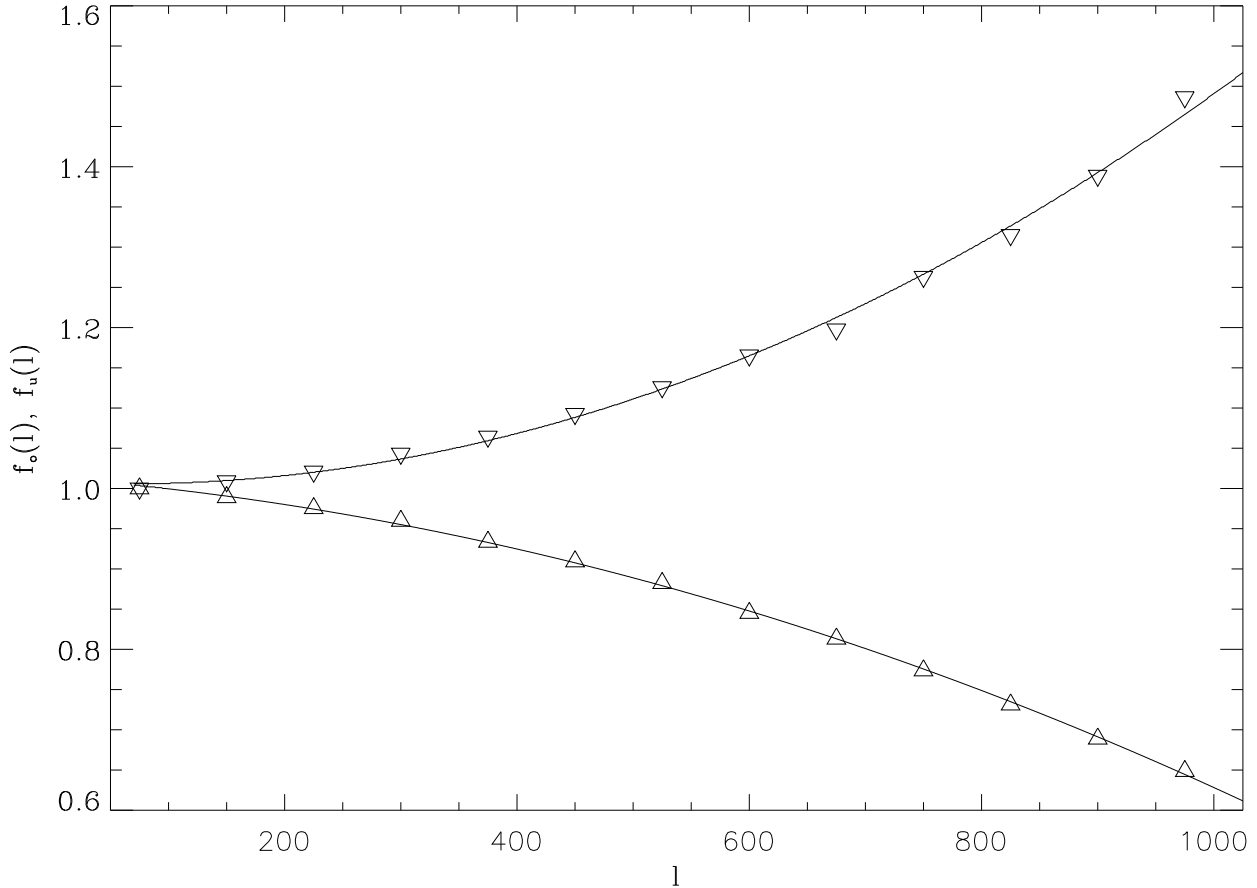}
\caption{The correction factors $f_{o}(\ell)$ (upward pointing triangles)
and $f_{u}(\ell)$ (downward pointing triangles) for the correlated CMB 
power spectrum error caused by the uncertainty of the effective beam 
width of the Boomerang experiment as given in Netterfield \etal~ (2001).
The solid lines are the fitting functions $f_o$ and $f_u$ given in the 
text.}
\label{beam}
\end{figure}

We estimate the beam width uncertainty as follows:
Using the data of Netterfield \etal~ (2001) for the $1\sigma$ dispersion 
of the CMB 
power spectrum due to beam width uncertainty, we have to estimate 
its effect on determination of peak and dip locations and amplitudes. To 
take into account the effect of  a $1\sigma$ overestimated beam 
width we have lowered the central points of the CMB power spectrum 
presented in Table~3  of \cite{boom01} by multiplying them by the 
$\ell$--dependent factor 
$$f_{o}(\ell)=1-1.1326\cdot 10^{-4}\ell-2.72\cdot 10^{-7}\ell^2.$$ 
To take into account the effect of a $1\sigma$ underestimated beam width 
we raise the central points of CMB power spectrum all by  multiplication
with the factor 
$$f_{u}(\ell)=1-6.99\cdot 10^{-5}\ell+5.53\cdot 10^{-7}\ell^2.$$  
($f_{o}(\ell)$ and  $f_{u}(\ell)$ are best fits to the 
Boomerang-2001 data and are shown in Fig.~\ref{beam}). For both cases we have 
repeated the peak and dip determination   procedure.
Best-fit values determined for the central points of CMB power spectrum
 give us the $1\sigma$ errors of the peak/dip characteristics due to beam 
width uncertainties. The results are presented in Table~\ref{tab_boom}. They 
show that error bars of all peak and dip {\em locations} caused by the 
beam width uncertainty are substantially less than the statistical 
errors. But they dominate for the {\em amplitude} of  the 3rd 
peak and are comparable with the statistical error for the amplitudes
of the 1st dip, the 2nd peak and the 2nd dip. However, the beam size
errors are significantly smaller than  statistical errors for the 1st 
acoustic peak. 

Since all sources of errors have different nature and are statistically 
independent they add in quadrature. The resulting symmetrized total errors
are shown in the before last and last columns of  
Table~\ref{tab_boom}. They are used in the cosmological parameter search 
procedure described in Section 4.

\begin{deluxetable}{crrrr}
\tabletypesize{\scriptsize}
\tablecaption{Best fit values  for locations ($\ell_p$) and amplitudes 
($A_p$, $[\mu K^2]$) of the peaks and dips in the CMB temperature 
fluctuation power spectrum measured by Boomerang 
(Netterfield et al. 2001). Statistical errors (1st upper/lower values)  
and errors caused by beam width uncertainties (2nd upper/lower values) 
are shown in columns 2 and 3.  The 20\% calibration uncertainty is 
included in the symmetrized total errors presented in the last column. 
\label{tab_boom}}
\tablewidth{0pt}
\tablehead{
\colhead{Features} & \colhead{$\ell_p$}&\colhead{$A_p$}&
	\colhead{ $\ell_p$}&\colhead{$A_p$} \\ [4pt] }
\startdata
1st peak &$212^{+13+2}_{-20-3}$  &$5426^{+540+112}_{-539-135}$&$212\pm 17$ &$5426\pm 1218$\\[4pt] 
1st dip  &$413^{+54+6}_{-27-6}$      &$1960^{+272+142}_{-282-158}$&$413\pm 50$ &$1960\pm 503$\\[4pt]
2nd peak &$544^{+56+14}_{-52-14}$    &$2266^{+275+309}_{-274-283}$&$544\pm 56$ &$2266\pm 607$\\[4pt]
2nd dip  &$746^{+114+9}_{-63-9}$     &$1605^{+373+422}_{-436-362}$&$746\pm 89$ &$1605\pm 650$\\[6pt]
3rd peak &$843^{+26+5}_{-42-7}$      &$2077^{+426+720}_{-411-573}$&$843\pm 35$ &$2077\pm 876$\\[4pt]
\enddata
\end{deluxetable}

\subsection{Adding DASI and MAXIMA-1 data}

\begin{figure}
\plotone{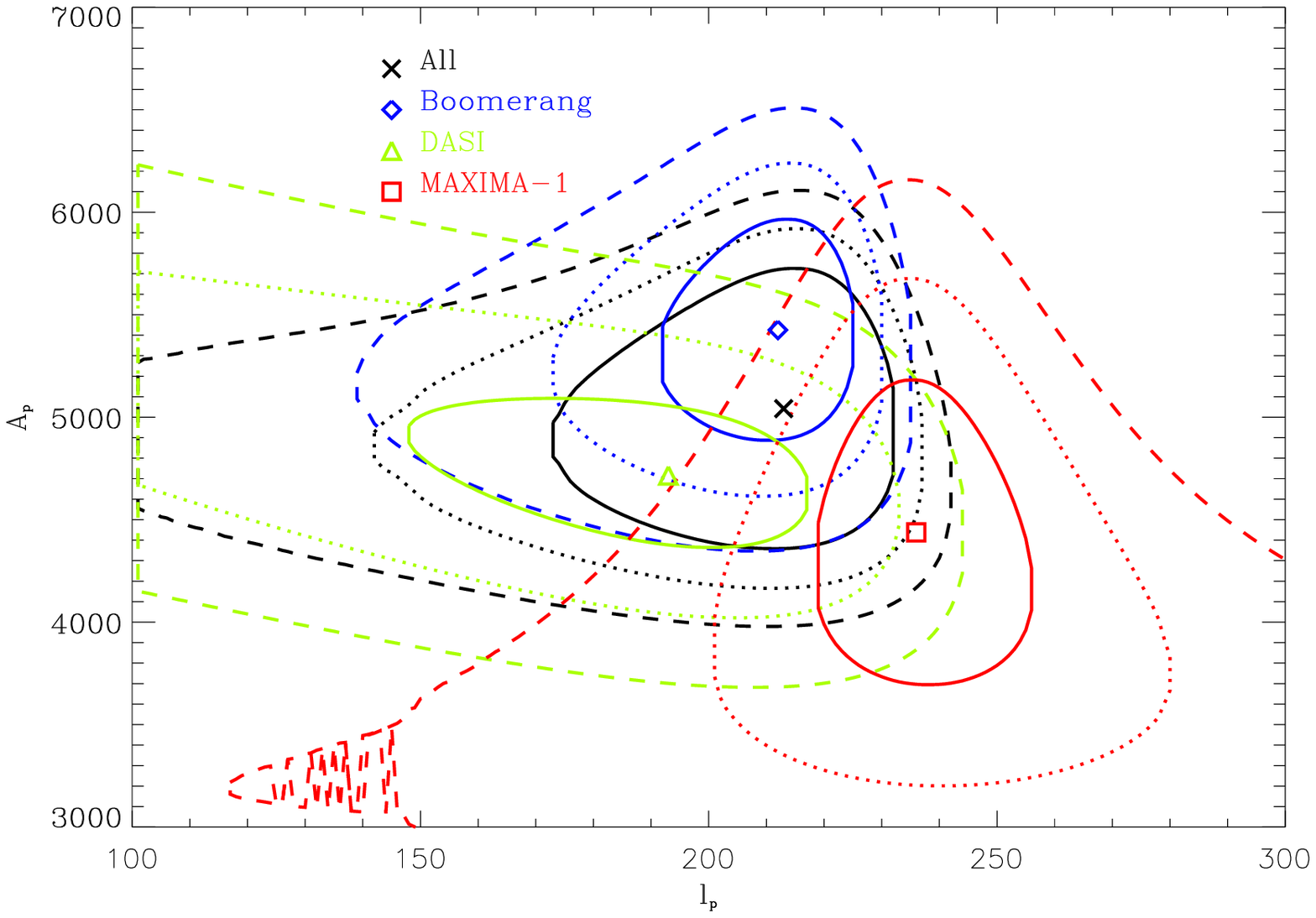}
\caption{The location of the first acoustic peak in the plane 
$\left(\ell, \ell(\ell+1)C_{\ell}/2\pi\right)$ for the Boomerang-2001 
(blue diamond), 
DASI (green triangle), MAXIMA-1 (red square) data and for all experiments
together (black cross) determined as maxima of corresponding best-fit 
parabola. The $1,\;2,\;3\sigma$ confidence contours are also shown. 
$\chi^2_{min}$, $N_f$ and $\Delta\chi^2$ for the first peak of the 
Boomerang-2001 data are given in the caption of Fig.~\ref{peaks}, for 
the other cases we have:  DASI -- $\chi^2_{min}=1.8$, $N_f=1$ and 
$\Delta\chi^2=1,\;4,\;9$ for the 
$1,\;2,\;3\sigma$ confidence contours accordingly, MAXIMA-1 -- 
$\chi^2_{min}=4.22$, $N_f=2$ and $\Delta\chi^2=2.3,\;6.17,\;11.8$, all 
experiments together --  $\chi^2_{min}=15.3$, $N_f=9$ and 
$\Delta\chi^2=10.43,\;17.18,\;25.26$. The dominant contribution 
to $\chi^2_{min}$ comes from the MAXIMA-1 data.}
\label{all1}
\end{figure}

\begin{figure}
\plotone{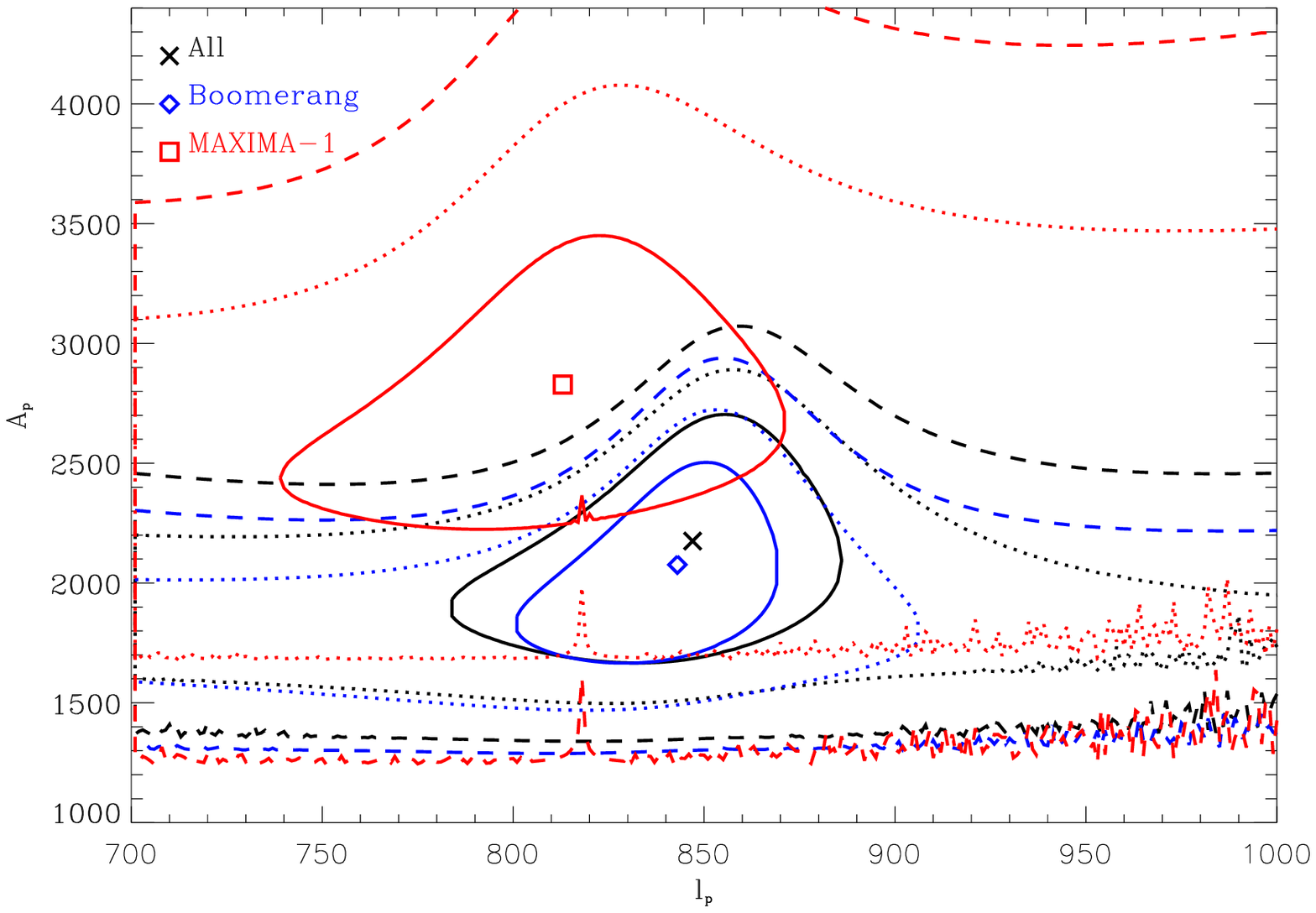}
\caption{The location of the third acoustic peak in the plane  $\left(\ell, 
\ell(\ell+1)C_{\ell}/2\pi\right)$ for the Boomerang-2001 (blue diamond) 
and MAXIMA-1 (red square) data and for both experiments together 
(black cross), determined as maxima of the corresponding best-fit 
parabola. The $1,\;2,\;3\sigma$ confidence contours 
are also shown. $\chi^2_{min}$, $N_f$ and $\Delta\chi^2$ for the 
Boomerang-2001 third acoustic peak is given in the caption of 
Fig.~\ref{peaks}, for the other cases we have: MAXIMA-1 -- 
$\chi^2_{min}=0.67$, $N_f=1$ and $\Delta\chi^2=1,\;4,\;9$, all 
experiments together --  $\chi^2_{min}=3.0$, $N_f=6$ and 
$\Delta\chi^2=7.04,\;12.82, \;20.06$.  The dominant contribution to 
$\chi^2_{min}$ comes from the MAXIMA-1 data.}
\label{all3}
\end{figure}

\begin{figure}
\plotone{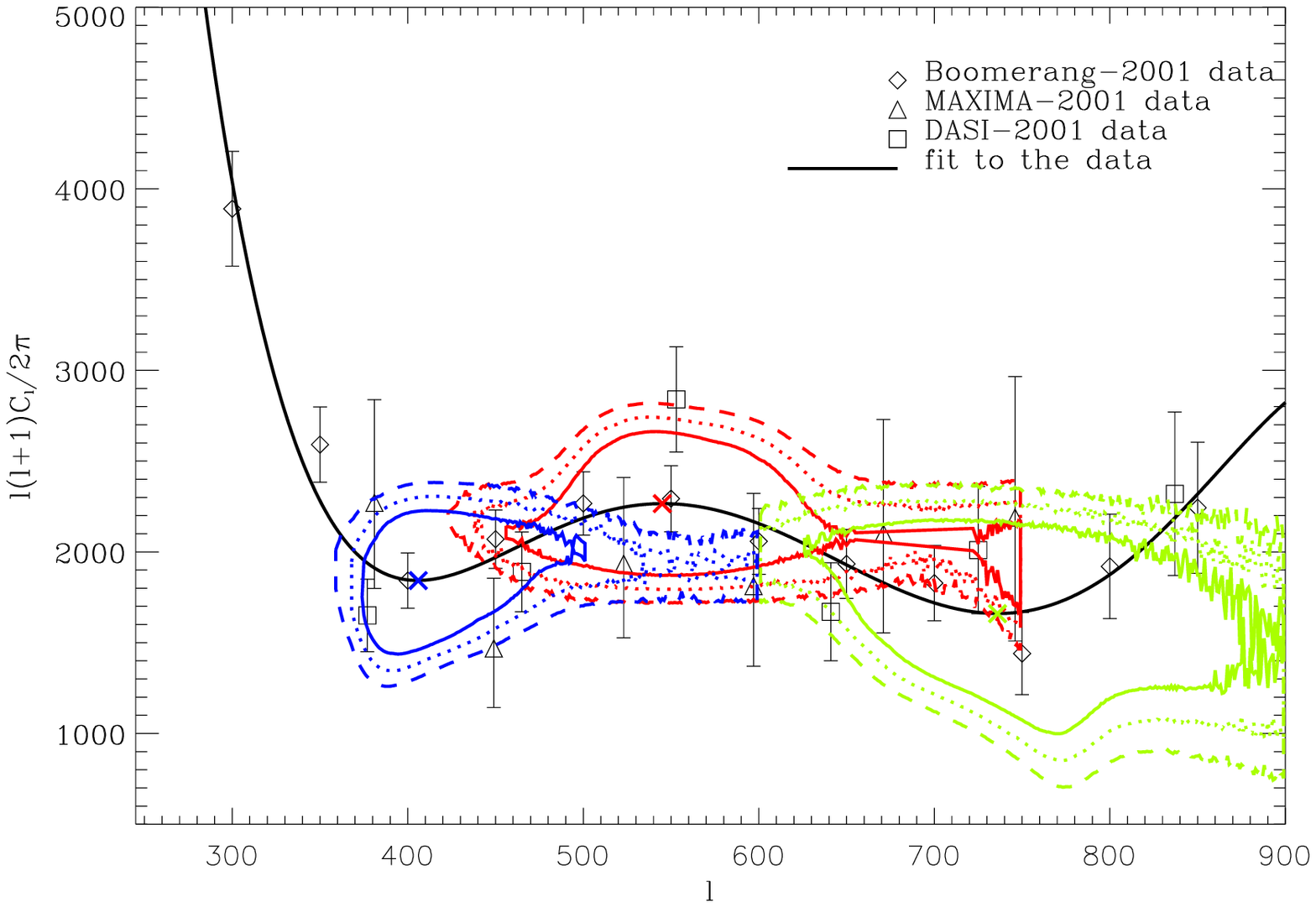}
\caption{The locations of the second acoustic peak and the two dips in the  
$\left(\ell,\ell(\ell+1)C_{\ell}/2\pi\right)$ plane are shown together 
with the  best-fit 5-th order polynomial. 
All points in this range from Boomerang, MAXIMA-1 and DASI have been 
used jointly. The second peak and the dips are determined as 
extrema of negative and positive curvature of the corresponding best fit
polynomial. The $1,\;2,\;3\sigma$ confidence contours are also shown. 
For all experiments together, the best-fit gives $\chi^2_{min}=17.8$ for $
N_f=18$ degrees of freedom.}
\label{all2}
\end{figure}

We have repeated the determination of peak and dip locations and amplitudes 
with the data of two other experiments, DASI (Halverson et al. 2001) and  
MAXIMA-1 \cite{maxima01}, released simultaneously with Boomerang-2001. Both 
confirm the main features of  the Boomerang CMB power spectrum: a 
dominant first acoustic peak at $\ell\sim 200$, DASI shows a second peak 
at $\ell\sim 540$ and MAXIMA-1 exhibits mainly a 'third peak' at 
$\ell\sim 840$. The results presented in Figs.~\ref{all1},\ref{all3} 
and \ref{all2} are quantitative figures of merit for their mutual 
agreement and/or  disagreement. In Fig.~\ref{all1} the 
$1,\;2\;{\rm and}\;3\sigma$ contours for the first peak location and 
amplitude for each experiment as well as the contours  
for the combined data are presented. 

In Fig.~\ref{all1} one sees that MAXIMA, like Boomerang,
indicates on the existence of the first acoustic peak at 
approximately 3$\sigma$ C.L. But its $1\sigma$
contour for location of this peak  in the  $\left(\ell, 
\ell(\ell+1)C_{\ell}/2\pi\right)$ plane does not intersect the Boomerang 
$1\sigma$ contour, though their projections on 
$\ell$ and $\ell(\ell+1)C_{\ell}/2\pi$ axes do. This can be caused by 
systematic (normalization) errors inherent in both experiments. 
Approximately a quarter of the area outlined by the Boomerang $2\sigma$ 
contour falls within the MAXIMA $2\sigma$ contour.
The experiments show the same level of agreement in the data on 3rd 
acoustic peak (Fig. \ref{all3}). In the range of the 1st dip - 2nd peak 
- 2nd dip, the MAXIMA data have no significant extrema,
even $1\sigma$ contours are open in both directions of the $\ell$ axis.

The DASI experiment establishes the location and amplitude of the first 
acoustic peak at somewhat more than $1\sigma$ but less than $2\sigma$. 
The remarkable feature is the intersection of the $1\sigma$ contours of 
DASI and Boomerang. 
Approximately 1/5 of the area outlined by the MAXIMA $2\sigma$ contour 
is within the corresponding DASI contour. 

Our analysis has also shown that the DASI data on the second acoustic  
peak agree very well with Boomerang;  the $1\sigma$ contours 
nearly superimpose. The agreement of these two experiments is 
impressive.

We have repeated the 
determination of peak and dip locations and amplitudes using the data of all  
experiments jointly. The contours for the combined data are shown by the 
thicker black lines in Figs. \ref{all1} and \ref{all3}. The C.L. 
contours for the 1st dip, 2nd peak and 2nd dip determined as 
regions of locations  of negative and positive extrema of a 5-th order 
polynomial fit are shown 
in Fig.~\ref{all2}. The comparison with the corresponding figure for  
the Boomerang data alone (Fig. \ref{dips}) shows their agreement.

The best-fit values of $\ell_{p_i}$, $A_{p_i}$ 
($i=1,\; 2,\; 3$) and  $\ell_{d_k}$, $A_{d_k}$ ($k=1,\; 2$) as well 
as their $1\sigma$  statistical errors are given in
Table~\ref{tab_all}.

Other methods of model-independent determinations 
of acoustic oscillation extrema were proposed by \cite{df01,mn01}.
Their analysis like ours finds low statistical
significance (less than $2\sigma$) for the detection of second and
 third peaks.

\begin{deluxetable}{crrrrrr}
\tabletypesize{\scriptsize}
\tablecaption{Best fit values for the locations and amplitudes 
of peaks and dips in the CMB temperature fluctuation power spectrum 
from the DASI and MAXIMA-1 experiments. Statistical  errors are 
determined as described in the text. In the last column the results 
obtained from the data of all three experiments together are presented. 
\label{tab_all}}
\tablewidth{0pt}
\tablehead{
\colhead{}&\multicolumn{2}{c}{DASI}&\multicolumn{2}{c}{MAXIMA-1}
&\multicolumn{2}{c}{All three experiments}\\  
\cline{2-3} \cline{4-5} \cline{6-7}\\
\colhead{Features} & \colhead{$ \ell_p$}&\colhead{$A_p$}&
\colhead{ $ \ell_p$}&\colhead{$A_p$}&\colhead{ $ \ell_p$}&
\colhead{$A_p$}\\ [4pt]
}
\startdata
1st peak  &$193^{+24}_{-45}$ &$4716^{+376}_{-351}$ 
&$236^{+20}_{-17}$ & $4438^{+743}_{-743}$ 
&$213^{+35}_{-59}$ &$5041^{+1017}_{-1196}$ \\[4pt] 
1st dip \tablenotemark{a}
&$378^{+15}_{-11}$ &$1578^{+170}_{-178}$ 
&$475^{+264}_{-83}$ &$1596^{+427}_{-443}$ 
&$406^{+97}_{-32}$ &$1843^{+385}_{-405}$ \\[4pt]
2nd peak  \tablenotemark{a}
&$536^{+30}_{-24}$ &$2362^{+176}_{-176}$ 
&$435-739$\tablenotemark{b} &$1500-2800$ \tablenotemark{b}
&$545^{+204}_{-89}$ &$2266^{+397}_{-609}$ \\[4pt]
2nd dip  \tablenotemark{a} 
&$709^{+45}_{-46}$ &$1799^{+221}_{-308}$ 
&$435-739$ \tablenotemark{b}&$1000-2700$ \tablenotemark{b}
&$736^{+163}_{-117}$ &$1661^{+517}_{-663}$ \\[4pt]
3rd peak  & -- & -- & $813^{+286}_{-112}$ &$2828^{+1880}_{-1584}$ 
&$847^{+252}_{-146}$ &$2175^{+897}_{-836}$ \\[4pt]
\enddata
\tablenotetext{a}{The extrema were determined by 
approximating the experimental CMB power spectrum by 5-th order polynomial}

\tablenotetext{b}{Just the ranges where the probability to find the peak or 
dip is $>68.3\%$ are indicated}
\end{deluxetable}

The peak locations and amplitudes  from the 
Boomerang-2001 CMB data presented in the  Table~\ref{tab_boom} 
show good quantitative agreement in the locations  and,  somewhat less
good, in the amplitudes obtained from the corresponding data of the 
other experiments and all the data together. The agreement can be 
improved when other error sources (calibration, beam width uncertainty, 
cosmic variance etc)  of each experiment are taken into account. With 
some luck, the new mission MAP, which has been launched successfully 
last June, will remove many of the current problems 
and will considerably improve the data on CMB power spectrum.

Clearly, the existence of the first peak in the spectrum is very well 
established in the present experiments. It has already been established
(with less accuracy) before (see, eg. \cite{kp00,nd00}).
Finally we have also assessed approximately the probability for each of the 
experiments to show no secondary peaks structure whatsoever. For the Boomerang
data this probability is less than 4.6\%, while for MAXIMA it is on the order 
of 15\% and for DASI about 8\%.

\section{Analytic determinations of  the locations and  amplitudes of 
the acoustic peaks and dips}

In order to use the data in Tables~\ref{tab_boom} and~\ref{tab_all} to 
determine cosmological parameters, we need a fast  algorithm 
to calculate the peak and dip positions for a given model. Here we 
improve the analytical approximations of peak/dip positions 
and amplitudes which have been derived in several papers 
\cite{efs99,hu01,dn01,dl01}.  We start by discussing the normalization 
procedure.

\subsection{Normalization of the density power spectrum}

 The 4-year COBE data, which establish the amplitude and the form of the 
CMB power spectrum at the largest angular scales ($\ell\le 20$),  are 
taken into account via the approximation for  $C_{10}$ 
proposed by~Bunn \& White (1997). This requires accurate calculations of 
$C_\ell$ in the range $\ell\le 12$. The dominant contribution 
on these angular scales is given by ordinary Sachs-Wolfe (SW) effect. 
However, the Doppler (D) effect and the cross-correlation term Sachs-Wolfe -- 
adiabatic (SW-A) in the general expression for the correlation 
function 
$\langle{\Delta T\over T}({\bf n_1})\cdot {\Delta T\over T}({\bf n_2})\rangle$ 
have to be taken into account as well if we want to achieve an accuracy 
better than 20\% (see Appendix A).  For $\Lambda$ dark matter models 
and models with non-zero 3-curvature, also the integrated Sachs-Wolfe 
effect (ISW) contributes. We use the  factors $K_\ell$ ($\ge 1$) 
introduced and calculated by Kofman \&~ Starobinsky (1985) and improved 
by Apunevych \&~ Novosyadlyj (2000), so that 
$C_\ell^{\rm SW+ISW}= K_\ell^2C_\ell^{\rm SW}$ (for details see 
Appendix~A). 

The normalization of the power spectrum of scalar perturbations then 
consist in two steps: 

i) We calculate 
\be 
C_\ell=C_\ell^{\rm SW+ISW}+C_\ell^{\rm D}+C_\ell^{\rm A}+C_\ell^{\rm SW-A} 
\label{f_cl}
\ee
(for $\ell=2,\;3,\;5,\;7,\;10,\;11,\;12$) by the analytical formulae 
given in Appendix~A with arbitrary normalization.  
This determines the  shape of the CMB power spectrum in the range of
the COBE data, and hence the best-fit parameter $C_{10}^{\rm COBE}$ 
to 4-year COBE and the first and second derivatives as defined in
Bunn \& White (1997) for models with given cosmological parameters;

ii) Since each term in the expression (\ref{f_cl}) is 
$\propto \delta_h^2$, 
where $\delta_h$ is the present matter density perturbations at horizon 
scale, we can now determine $\delta_h$ and along with it the  value of 
the normalization constant for scalar perturbations 
$A_s=2\pi^{2}\delta_{h}^{2}(3000{\rm Mpc}/h)^{3+n_s}$ for a model with 
given cosmological parameters. Here $n_s$ is the spectral index for 
primordial scalar density perturbations 
and $h$ is dimensionless Hubble parameter (in units of 100 km/sec/Mpc). 

Both these steps are also performed in CMBfast. Hence, our 
normalization procedure  for the power spectrum is equivalent to
normalization with  CMBfast. Calculations show that our  value
$C_{10}^{\rm COBE}$ never differs from the result of CMBfast by more 
than $3\%$. The accuracy of  the overall normalization constant 
$\delta_h$ for $\Lambda$DM models with appropriate values of 
parameters is better then $5\%$. This has been controlled by comparing 
the value of  $\sigma_8$  from  CMBfast with our semi-analytical 
approach. This error simply reflects the accuracy of the analytical 
approximation of the transfer function for 
density fluctuations by Eisenstein \&~ Hu (1999) which we have used.

\subsection{Positions and  amplitudes of CMB extrema: analytic approach}

One of the main ingredients for our search procedure is a fast and 
accurate calculation of the positions and amplitudes of the acoustic 
peaks and dips, which depend on cosmological parameters.

The dependence of the position and amplitude of the first acoustic
peak of the CMB power spectrum on cosmological
parameters has been investigated using CMBfast.
As expected, the results are, within reasonable
accuracy, independent of the hot dark matter
contribution ($\Om_{\nu}$). This was also shown by Novosyadlyj \etal~ 
(2000). For the remaining parameters, $n_s$, $h$, $\Omega_b$, 
$\Omega_{cdm}$ and $\Omega_{\Lambda}$, we determine the resulting 
values $\ell_{p_1}$ and $A_{p_1}$ using the analytical approximation 
given by Efstathiou \&~ Bond (1999) and Durrer \&~ Novosyadlyj (2001).
In these papers the CMB anisotropy spectrum is approximated in the 
vicinity of the first acoustic peak by
\bea
\nonumber 
{\ell(\ell+1)\over 2\pi}C_{\ell}={\ell(\ell+1)\over 2\pi}(C^{SW}_{\ell}+0.838C^{SW}_2\cdot\\
{\cal A}(\Omega_b,\Omega_{cdm},\Omega_k,n_s,h)\exp
 \left[ -{(\ell-\ell_{p_1})^2\over 2(\Delta\ell_{p_1})^2}\right])~,  
\eea
where $\Delta\ell_{p_1}=0.42\ell_{p_1}$, $C^{SW}_{\ell}$ is the Sachs-Wolfe 
approximation for the $C_{\ell}$s derived in the Appendix ~A, 
Eq.~(A9), and 
\bea
\nonumber
{\cal A}(\Omega_b,\Omega_{cdm},\Omega_k,n_s,h) = \\ 
\nonumber
\exp{[a_1+a_2\omega_{cdm}^2+a_3\omega_{cdm}+a_4\omega_b^2+a_5\omega_b+}\\
{+a_6\omega_b\omega_{cdm}+a_7\omega_k + a_8\omega^2_k+a_9(n_s-1)]}. 
\eea
Here
$\omega_b\equiv\Omega_bh^2$, $\omega_{cdm}\equiv\Omega_{cdm}h^2$,
$\omega_k\equiv(1-\Omega_m-\Omega_{\Lambda})h^2$. 
The position of the acoustic peaks is determined as in \cite{efs99} for 
open and flat models and in \cite{dn01} for closed models.
The coefficients $a_i$ are defined  by fitting to the numerical CMBfast 
amplitudes of the first acoustic peak on a sufficiently wide grid of 
parameters. We find: $a_1=2.503$, $a_2=8.906$, $a_3=-7.733$, 
$a_4=-115.6$, $a_5=35.66$, $a_6=-7.225$, $a_7=1.96$, $a_8=-11.16$, 
$a_9=4.439$. The accuracy of the approximation is better than 5\% in 
the parameter range $0.2\le\Omega_m\le 1.2$, 
$0\le\Omega_{\Lambda}\le 0.8$, $0.015\le \Omega_b\le 0.12$, 
$0.8\le n_s\le 1.2$ and $0.4\le h\le 1.0$. The approximation for the 
amplitude breaks down in the models with large curvature 
($\Omega_k\le -0.2$ and $\Omega_k\ge 0.6$) and low baryon density, ($\omega_b 
 \ll 0.006$).

To calculate the  amplitudes of the 2nd and 3rd peaks, we use the 
analytic relations for the relative heights of these peaks w.r.t
the first peak as given by Hu et al. (2001).
\bea
\nonumber
A_{p_2}=A_{p_1}H_2(\Omega_m,\Omega_b,n_s), \\
A_{p_3}=A_{p_1}H_3(\Omega_m,\Omega_b,n_s),
\eea
where the functions $H_2$ and $H_3$ are given in Eqs. (B16) and (B17) 
respectively. For the locations of 2nd and 3rd peaks we 
 use the analytic approximations  given by Hu \etal~ (2001)
and Doran \& Lilley (2001) (see also Appendix B).

Unfortunately, we have no analytic approximation for the dip amplitudes 
 and hence we can not use their experimental values to determine
cosmological parameters. But a sufficiently accurate analytic 
approximation for the location of the 1st dip is given in Doran \&~ Lilley 
(2001). We use it here.

Hence, we have analytical approximations for the dependences of the 
positions and amplitudes of three acoustic peaks and the location of the 
1st dip on cosmological parameters
\bea
&\ell_{p_i}(\Omega_m,\Omega_{\Lambda},\Omega_k,\Omega_b,h),&\nonumber \\
&A_{p_i}(\Omega_m,\Omega_{\Lambda},\Omega_k,\Omega_b,n_s,h),&\;\;\; (i=1,2,3) \nonumber \\
&\ell_{d_1}(\Omega_m,\Omega_{\Lambda},\Omega_k,\Omega_b,h).& \nonumber 
\eea
Comparing the analytical values for different sets of parameters
with numerical calculations using CMBfast, shows that the accuracy is 
about 5\% for all locations 
and amplitudes of 1st and 3rd peaks in the ranges of cosmological parameters
indicated above.
The accuracy for amplitude of the 2nd peak is always 
better than $9$\% in the same ranges.
For some parameter values the second peak is underestimated.

Of course, in principle the half-widths of peaks and dips, their "smoothness",
"sharpness" or their inflection points may contain additional information 
on cosmological parameters. But when one compares CMB spectra obtained in the
different experiments up to date, one concludes that their accuracy is not 
sufficient to influence the resulting cosmological parameters  at the
present precision. Therefore, 
we only use the most prominent observable patterns of the CMB power spectrum - 
locations and amplitudes of acoustic peaks.

For the convenience, we present all analytic approximations 
used here  in Appendix B.

\section{Cosmological parameters from the CMB peak amplitudes and 
locations}

 We now use the results of Sections 2 and 3 to determine the cosmological
parameters $\Omega_m$, $\Omega_{\Lambda}$, $\Omega_{\nu}$ (one sort of 
massive neutrino), $\Omega_b$, $n_s$, $h$, $T/S$ 
($\equiv C^{tensor}_{10}/C^{scalar}_{10}$) and $\tau_c$ 
(optical depth to decoupling). We use the method described in detail in a
previous paper  \cite{dn01}. We include 8 experimental points from the CMB 
power spectrum (COBE $C_{10}$, the amplitudes and locations of three acoustic 
peaks and the location of the first dip). In order to have a positive number
of degrees of freedom, $N_f\ge 1$ we add a weak constraint for the Hubble 
constant, $0.5\le h \le 0.8$ when searching for 8 parameters.

In order to establish $1\sigma$ confidence intervals for each parameter we 
have applied the marginalization procedure described in  
(Durrer \& Novosyadlyj 2001). The results are presented in 
Table~\ref{tab_par}.

At first we check how the best fit parameters depend on the accuracy of 
the  peak/dip locations and amplitudes.
For this we compare the resulting  cosmological parameters from the data 
given in Table \ref{tab_boom} if we consider only statistical errors 
and with total errors. The results are given in the first two rows of 
Table~\ref{tab_par}. In spite of the different errors of the experimental 
values and their relations (the total errors of peak amplitudes increase 
faster with the peak number than the statistical error) the best fit
 cosmological parameters are similar. 

In order to estimate the sensitivity of cosmological parameters to experimental
values we substitute the Boomerang data on peak/dip locations and amplitudes 
by the values from Table~\ref{tab_all}, obtained from all three experiments
combined. The comparison of the results in the third row with those above 
shows that the results are practically unchanged. We believe that the
Boomerang data on peak/dip locations and amplitudes are best studied and 
have well established statistics, hence we use them in the following 
determinations.

The neutrino contents for these three data sets can be rather large, 
$\Omega_\nu \sim 0.3$. This is due to the low sensitivity of 
the CMB anisotropy spectrum to $\Omega_{\nu}$. When 
$\Omega_{\nu}=0$ is fixed, the best-fit values for the remaining parameters
stay practically unchanged and also $\chi^2$ increases only very little.
The CMB doesn't care whether dark matter should be hot or cold. In order to 
distinguish between cold and hot dark matter data which is
sensitive to the density power spectrum on smaller scales needs to be 
added.  In Table~\ref{tab_par} we therefore exclude $\Omega_{\nu}$ from 
the determination procedure and fix its value to $0$ in the first five rows. 
Contrary, for $T/S$ and $\tau_c$ we obtain $0$, but with large 
$1\sigma$ confidence limits due to the degeneracy in $T/S$, $\tau_c$ and $n_s$ 
\cite{efs99}.

A remarkable result is the good agreement of the best-fit content of
baryons $\Omega_bh^2\approx 0.02$ with the constraint from standard 
nucleosynthesis  and the observed intergalactic content of light 
elements \cite{burles}.
The large $1\sigma$ confidence limits for $\Omega_m$ and $\Omega_{\Lambda}$ are due to
the well known degeneracy of the CMB power spectrum in these 
parameters \cite{efs99}. However, the sum of their best fit values, 
$\Omega_m+\Omega_\Lambda = 1 -\Omega_k$ is always very close to $1$ which
implies that spatial curvature is small for the best fit model.

\begin{deluxetable}{crrrrrrrrr} 
\tabletypesize{\scriptsize}
\tablecaption{Cosmological parameters from the extrema of the CMB angular 
power spectrum in combination with other cosmological data sets.
The upper/low values show $1\sigma$ confidence limits which are obtained by maximizing
the (Gaussian) 68 percent confidence contours over all other parameters. 
The LSS data set is the same as in \cite{dn01}. \label{tab_par}}
\tablewidth{0pt}
\tablehead{
\colhead{Observable data set} & \colhead{$\chi^2_{min}/N_f$}&
\colhead{$\Omega_{\Lambda}$}&\colhead{$\Omega_m$}&\colhead{$\Omega_{\nu}$}
&\colhead{$\Omega_b$}&\colhead{$n_s$}&\colhead{$h$}&\colhead{$T/S$}
&\colhead{$\tau_c$} \\ [4pt] }
\startdata
CMB$_{(Boom, stat.)}$&1.01/2&$0.69^{+0.23}_{-0.56}$&$0.31^{+0.61}_{-0.21}$&0$^{*)}$&$0.055^{+0.13}_{-0.028}$&$0.89^{+0.81}_{-0.08}$&$0.65^{+0.23}_{-0.24}$&0$^{+27}$&0$^{+1.65}$ \\[4pt] 
CMB$_{(Boom, total)}$&0.95/2&$0.64^{+0.31}_{-1.42}$&$0.36^{+1.04}_{-0.35}$&0$^{*)}$&$0.057^{+0.18}_{-0.047}$&$0.89^{+0.97}_{-0.14}$&$0.65^{+0.23}_{-0.23}$&0$^{+44}$&0$^{+1.90}$ \\[4pt] 
CMB$_{(All, stat.)} $&0.09/2&$0.63^{+0.35}_{-1.35}$&$0.37^{+1.04}_{-0.36}$&0$^{*)}$&$0.051^{+0.29}_{-0.05}$&$0.90^{+1.30}_{-0.11}$&$0.65^{+0.23}_{-0.24}$&0$^{+20}$&0$^{+1.75}$ \\[4pt] 
CMB$_{(Boom, total)}$+&&&&&&&& \\[4pt] 
$h$ $\&$ BBN\tablenotemark{a} &1.11/3&$0.69^{+0.26}_{-1.30}$&$0.31^{+1.05}_{-0.24}$&0$^{*)}$&$0.047^{+0.048}_{-0.018}$&$0.90^{+0.56}_{-0.10}$&$0.65^{+0.20}_{-0.19}$&$0^{+2.7}$&$0^{+0.90}$\\[4pt]
$h$, BBN $\&$ SNIa\tablenotemark{b}  &1.11/4&$0.72^{+0.17}_{-0.21}$&$0.29^{+0.15}_{-0.13}$&0$^{*)}$&$0.047^{+0.048}_{-0.02}$&$0.90^{+0.60}_{-0.12}$&$0.65^{+0.22}_{-0.19}$&$0^{+3.5}$&$0^{+1.1}$\\[4pt]
$h$, BBN $\&$ LSS\tablenotemark{c}   &8.22/11&$0.46^{+0.31}_{-0.46}$&$0.48^{+0.52}_{-0.22}$&$0.06^{+0.20}_{-0.06}$&$0.047^{+0.12}_{-0.026}$&$1.03^{+0.59}_{-0.23}$&$0.66^{+0.31}_{-0.31}$&$0^{+3.5}$&$0.15^{+0.95}_{-0.15}$\\[4pt]   
$h$, BBN, SNIa $\&$ LSS\tablenotemark{d}&10.4/12&$0.64^{+0.14}_{-0.27}$&$0.36^{+0.21}_{-0.11}$&$0.00^{+0.17}$&$0.047^{+0.093}_{-0.024}$&$1.0^{+0.59}_{-0.17}$&$0.65^{+0.35}_{-0.27}$&$0^{+1.7}$&$0.15^{+0.95}_{-0.15}$\\[4pt]
the same\tablenotemark{e}&11.6/14&$0.61^{+0.16}_{-0.26}$&$0.37^{+0.21}_{-0.13}$&$0.00^{+0.11}$&$0.041^{+0.043}_{-0.023}$&$0.95^{+0.17}_{-0.14}$&$0.70^{+0.34}_{-0.20}$&0$^{*)}$&0$^{*)}$ \\[4pt]  \enddata

\tablenotetext{*)} {This parameter is fixed to $0$.} 
\tablenotetext{a} {The big bang nucleosynthesis constraint on baryon content  $\widetilde{\Omega_bh^2}=0.02\pm 0.001$ from Burles \etal~(2001) is included.}

\tablenotetext{b} {The constraint on the $\Omega_{\Lambda}-\Omega_m$ 
relation  from SNIa distance measurements \cite{per99}, 
$\widetilde{[\Omega_m-0.75\Omega_{\Lambda}]}=-0.25\pm 0.125$ is added.}

\tablenotetext{c} {In addition to the parameters given in the different 
columns, we have also to determine the Abell-ACO biasing parameter, $b_{cl}$.
The result is: $b_{cl}=2.64\pm 0.27$} 

\tablenotetext{d} {For this data set we obtain $b_{cl}=2.47\pm 0.19$}
\tablenotetext{e} {$b_{cl}=2.5\pm 0.2$}

\end{deluxetable}

We repeat the search procedure for different combinations of the CMB power 
spectrum extrema data (Table \ref{tab_boom}) with other cosmological 
data sets.  The LSS data set used here ranges from the Lyman alpha 
forest, determining  amplitude and spectral index of the matter power 
spectrum at very small scales, to large scale bulk velocities, cluster 
abundances and Abell cluster catalogs which determine $\sigma_8$ and 
the position of the 'knee' in the matter power spectrum. All of this is 
extensively discussed in \cite{dn01}.

The results for cosmological parameters from different combinations
of observational data  are shown in the lines 4 to 8 of Table~\ref{tab_par}.

Adding a stronger constraint on the Hubble parameter, $h =0.65\pm 0.10$, 
and the big bang nucleosynthesis (BBN) constraint 
changes the best fit cosmological parameters only slightly. The 
SNIa constraint on the relation between $\Omega_{\Lambda}$ and $\Omega_m$  
\cite{per99}  substantially reduces the errors 
of these parameters (5th line in Table \ref{tab_par}) as
 it removes the degeneracy between them. This degeneracy is also removed 
when we combine  CMB and LSS data. 

The  cosmological parameters obtained from the  Boomerang CMB power 
spectrum extrema data combined with all other cosmological measurements 
(a detailed list can be found in Durrer \& Novosyadlyj 2001)
are presented in lines 6th, 7th  and 8th of Table~\ref{tab_par}. The best fit 
values for the tensor mode amplitude $T/S$ defined as $C^T_{10}/C^S_{10}$ 
in the last two cases are practically zero but the $1\sigma$ confidence 
limits are wide due to the degeneracy of the CMB extrema  in $n_s,~T/S$ and 
$\tau_c$. Even when combining the  CMB with LSS data, the degeneracy in 
$n_s$ and $T/S$ is not significantly removed.  This is so, since a blue 
spectrum, which allows for a high tensor contribution to the CMB, can be 
compensated with a neutrino component which leads to damping of the matter 
power spectrum on small scales. If massive neutrinos are not allowed, the 
degeneracy between  $n_s$ and $T/S$ is lifted as soon as small scale LSS 
data is included. 

The best-fit values of spectral index 
$n_s$ in all cases are in the $1\sigma$ range of the value obtained from the
COBE 4-year data, $n_s= 1.2\pm 0.3$ \cite{ben96,cobe96}. 
When using the best fit model to calculate the data used to find it,
practically all  results are within the $1\sigma$ error range of the 
corresponding experimental data. Only two out of 31 experimental points are 
slightly outside. Namely the best-fit value of $\Omega_m-0.75\Omega_{\Lambda}$
in the last determination is at $1.1\sigma$ lower of its experimental value 
followed from SNIa test and $\sigma_8$ constraint established by \cite{bah98} 
from the existence of three massive clusters of galaxies is at $1.18\sigma$ 
higher than model predicted value. But the value of $\sigma_8$ in our
best-fit model, $\sigma_8=0.91$, is in the range of current estimates 
from  the Sloan Digital Sky Survey $\sigma_8^{(SLOAN)}=0.915\pm 0.06$, 
\cite{sdss01}), which is not included in our data set. 
The high degree of consistency within completely independent cosmological data 
sets is very encouraging. 

Moreover, all parameters of our best-fit model agree well with those 
extracted from the full Boomerang data \cite{boom01} combined with 
LSS and SNIa priors (compare our values 
derived from CMB+LSS in the 7th row of Table 3 and theirs in 
the 4th row of Table 4). But our 1$\sigma$ ranges for  
most parameters are significantly wider. That is due to two facts. First 
of all, we allow also for a tensor component and neutrinos. This increase in
the number of parameters also increases the degeneracies (e.g. between the 
tensor amplitude and the spectral index $n_s$) thereby enlarging the
errors of the physical parameters. In this sense our parameter errors
are rather to be compared with those of Wang et al. (2001) which allow 
for roughly the same degrees of freedom. But even their parameter estimation 
is somewhat more precise than ours. This is because 
we  use more conservative errors for the peak and dip 
locations and amplitudes which include 
statistical, normalization and beam uncertainties. 
We think that with present data, and with the model assumptions made by our 
choice of parameters, our precision is realistic. Clearly, future 
experiments like the MAP satellite will improve this situation.

Finally, to compare with the cosmological parameters obtained 
 in our previous  paper (Durrer \& Novosyadlyj 2001, Table 4), 
where the same LSS data set was used, we have repeated the search procedure
fixing $T/S =\tau_c=0$. The best-fit values of the parameters 
with $1\sigma$ errors obtained by maximizing the confidence contours over all 
other parameters are given in the last row of Table \ref{tab_par}. Comparing
them with values in the last column of Table 4 from \cite{dn01} shows that both
determinations have best-fit values in the $1\sigma$ confidence limits 
of each other. There the best fit model has a slight positive curvature, here 
a slightly negative. The $1\sigma$ confidence ranges here are somewhat 
wider than those obtained in the previous determination. These differences 
are due to the different CMB observable data set and the different 
normalization procedure. Even though in our previous analysis we have only 
taken into account the first peak. The errors in its location and amplitude
were significantly underestimated, leading to smaller error bars.

\section{Conclusions}

We have carried out a model-independent analysis of recent  CMB power
spectrum measurements in the Boomerang \cite{boom01}, DASI \cite{dasi} and 
MAXIMA (Lee et al. 2001) experiments and we have determined the locations and 
amplitudes of the first three acoustic peaks and two dips as well as their
confidential levels  (Table 1-2, Fig. 1-7).

In the Boomerang experiment the second and third acoustic peaks are 
determined at a confidence level somewhat higher than $1\sigma$.
Experimental errors which include statistics and systematics are 
still too large to establish the secondary peak locations and amplitudes 
at $2\sigma$ C.L. Only the position of one (the third ) secondary peak can 
be bounded from above $\ell_{p_3}\le 900$, at $2\sigma$ C.L. The same 
situation  is encountered when determining the locations and amplitudes of 
the first and second dips. However, the location and amplitude of the first 
peak, are well established  with confidence level, higher than $3\sigma$. 

The MAXIMA experiment also shows the existence of the first acoustic 
peak at approximately the same confidence level as Boomerang. 
But the $1\sigma$ contours for the peak position in the plane 
$\left(\ell, \ell(\ell+1)C_{\ell}/2\pi\right)$ do not intersect. However, 
their projections 
onto the $\ell$ and $\ell(\ell+1)C_{\ell}/2\pi$ axes do.  Approximately one 
quarter of the area inside the Boomerang $2\sigma$ contour falls within the
corresponding MAXIMA contour. The same level of agreement of 
these experiments is found in the data on the 3rd acoustic peak. In the 
range of 1st dip - 2nd peak - 2nd dip the MAXIMA data give no significant
information. Even the $1\sigma$ contours are open in both directions of the
$\ell$ axis.

The DASI experiment establishes the location and amplitude of the first 
acoustic peak at somewhat higher than $1\sigma$ C.L. but less than 
$2\sigma$. The $1\sigma$ contours for the position of the first peak
of the DASI and Boomerang experiments intersect. 
 Approximately 1/5 of  the area outlined by the MAXIMA $2\sigma$ contour is 
within the DASI $2\sigma$ contour. The DASI data on second acoustic peak
is in excellent agreement with the Boomerang results - the $1\sigma$
contours practically coincide.   

We have also determined the locations and amplitudes of the acoustic peaks 
and dips using the data of all three experiments. The results are very 
close to those from the Boomerang data alone.

To determine cosmological parameters from these data, we 
have improved the analytical approximations for the peak positions and 
amplitudes to an accuracy (determined by comparing the approximations with
the results of CMBfast) better than 5\%  in a sufficiently wide range
of parameters. 
% $0.2\le\Omega_m\le 1.2$, $0\le\Omega_{\Lambda}\le 0.8$,
% $0.015\le \Omega_b\le 0.12$, $0.8\le n_s\le 1.2$, $0.4\le h\le 0.9$. 
We have also developed a fast and accurate analytical method to
normalize the power spectrum to the 4-year COBE data
on $C_{10}$. Our analytical approximation is accurate to a few percent 
(in comparison to CMBfast) 
when all main effects (ordinary Sachs-Wolfe effect, integrated Sachs-Wolfe 
effect, adiabatic term, Doppler term and their mutual cross-correlations) 
are taken into account. For example, in the 
model with parameters presented in the last row of Table~\ref{tab_par} the 
relation of contribution from these components at $\ell=10$ are 
$C_{10}^{\rm SW}:C_{10}^{\rm A}:C_{10}^{\rm SW-A}:C_{10}^{\rm D}=1:0.098:-0.24:0.42$.

The cosmological parameters extracted from the data on locations and 
amplitudes of the first three peaks and the location of the first dip are in 
good agreement with other determinations (Netterfield et al. 2001; de 
Bernardis et al. 2001; Pryke et al. 2001; Wang et al. 2001; Durrer \& 
Novosyadlyj 2001).
That shows also that present CMB data can essentially be compressed into the 
height and slope of the Sachs-Wolfe plateau (at $\ell=10$) and the positions 
and amplitudes of the first three acoustic peaks and the first two dips.

A remarkable feature is the coincidence of the baryon content obtained from 
the CMB data, $\Omega_bh^2\approx 0.02$ with the value from 
standard nucleosynthesis ($0.02\pm 0.001$) \cite{burles}. Moreover, the 
CMB data together with constraints from direct measurements of the
Hubble constant, the SNIa data, the baryon content and the large scale 
structure of the Universe (the power spectrum of rich clusters, 
the cluster mass function, the peculiar velocity field of galaxies,
Ly-$\alpha$ absorption lines as seen in quasar spectra) select a best-fit 
model which gives predictions within about 1$\sigma$ error bars of all 
measurements. The cosmological parameters of this model are 
$\Omega_{\Lambda}=0.64^{+0.14}_{-0.27}$, $\Omega_{m}=0.36^{+0.21}_{-0.11}$,
$\Omega_b=0.047^{+0.083}_{-0.024}$, $n_s=1.0^{+0.59}_{-0.17}$, 
$h=0.65^{+0.35}_{-0.27}$ and $\tau_c=0.15^{+0.95}_{-0.15}$. The 
best-fit values of $\Omega_{\nu}$ and $T/S$ are close to zero, their
1$\sigma$ upper limits are $\Omega_{\nu}\le 0.17$, $T/S\le 1.7$. 

The cosmological parameters determined from the  CMB acoustic peak/dip 
locations and amplitudes data show good agreement with other cosmological
measurements and indicate the existence of a simple (adiabatic) best-fit 
model for all the discussed cosmological data  within the accuracy 
of present experiments.

\acknowledgments
It is a pleasure to acknowledge stimulating discussions with Alessandro 
Melchiorri, Roman Juszkiewicz and Roberto Trotta. 
BN is grateful to the Tomalla foundation for a
visiting grant and to Geneva University for hospitality.  RD thanks the 
Institute for Advanced Study for hospitality and acknowledges support 
from the Monell Foundation.

\appendix
\begin{center}
{\Large\bf  APPENDIX} \vspace{0.5cm}\\
\end{center}

\section{An analytic approximation for the CMB power spectrum at  large 
angular scales}
On sufficiently large angular scales (larger than the Silk damping scale) 
 temperature fluctuations in the CMB
can be related to density, velocity and metric perturbations at the 
last scattering surface and at later times by integrating the geodesic 
equation, similar to the classical paper by Sachs and Wolfe (1967).
Here we discuss only scalar perturbations. Tensor perturbations can  be
simply added  to the result and do not pose any significant difficulty.
Scalar perturbations generate CMB temperature fluctuations which can 
be written in gauge-invariant form as a sum of  four terms -- the 
ordinary Sachs-Wolfe effect, the integrated Sachs-Wolfe term, the Doppler term  
and the acoustic term \cite{d90}.  
\begin{eqnarray}
\left({\De T\over T}\right)^{(s)}(\eta_0,\bx_0,\bn) &=& 
{1\over 4}D_r(\eta_{dec}, \bx_{dec})+ 
 V_i(\eta_{dec}, \bx_{dec})n^i
 +(\Phi-\Psi)(\eta_{dec}, \bx_{dec}) \nonumber \\
 && -\int_{\eta_{dec}}^{\eta_0}
  (\Phi'-\Psi')(\eta,\bx(\eta))d\eta~.  \label{ani}
\end{eqnarray}
Here $\Phi$ and $\Psi$ are the Bardeen potentials \cite{bar80}, $V_i$ is 
the baryon velocity and $D_r$ is a gauge invariant variable for the radiation
density fluctuations. A prime denotes the partial derivative w.r.t. 
conformal time $\eta$. For  perfect fluids and for dust we have $\Psi=-\Phi$.
In Newtonian limit the Bardeen potentials just reduce to the ordinary 
Newtonian potential. For adiabatic perturbations 
${1\over 4}D_r = {1\over 3}\delta_m - {5\over 3}\Phi$ (see \eg~ 
Durrer \& Straumann (1999)),
where $\delta_m$ is the usual matter density perturbation., it
corresponds to  $\epsilon_m$ in Bardeen's notation. The variables 
$\eta$ and $\bx$ are conformal time and comoving position.

In realistic models, cosmological recombination  and decoupling of radiation 
from matter take place when 
$\rho_{m}>\rho_{r}$. Hence the large angular scale 
CMB power spectrum can be expressed in the terms of solutions of Einstein's 
equations for adiabatic linear perturbations in a dust Universe. 
The CMB anisotropies on angular scales $\theta\ge 10^o \;(\ell\le 20)$ 
are generated mainly by the  linear perturbations  of matter density, 
velocity and the gravitational potential at  scales much larger than the 
particle horizon at decoupling. Our approximation makes use of these facts.

We use the solutions of Einstein's equations for linear density perturbations 
in  flat models of a Universe with dust and a cosmological constant which can 
be found in \cite{ks85,an00}. The growing mode of density, velocity 
and gravitational potential perturbations, using the  gauge-invariant 
variables introduced by Bardeen (1980) and normalizing the scale factor 
$a(t_0)=a_0=1$, are
\be
\Phi(t,k)=K_{\delta}(t)C(k),~\delta_m(t,k)=-{2C(k)k^2a(t) K_{\delta}(t)\over 
3H_0^2\Omega_m},~
V^{\alpha}(t,k)=-i{2C(k)k^{\alpha} a(t)\dot a(t)K_V(t)\over 3H_0^2\Omega_m}.
\label{sol}
\ee
 $C(k)$ is (up to the time dependent factor $K_\delta$) the Fourier 
transform of the Bardeen potential, so that 
 $\Phi(t,{\bx})=(2\pi)^{-3/2}\int\Phi(t,\bk)e^{i{\bf kx}}d^3k$, 
$\delta_m(t,{\bf x})=(2\pi)^{-3/2}\int\delta(t,\bk)e^{i{\bf kx}}d^3k$ and  
$V^{\alpha}(t,{\bx})=(2\pi)^{-3/2}\int V^{\alpha}(t,\bk) e^{i{\bf kx}}d^3k$. 
The factors 
$K_{\delta}(t)\equiv {5\over 3}\left(1-\dot a/a^{2}\int_0^tadt \right)$ and
$K_{V}(t)\equiv{5\over 3}\left(\dot a/a^{2}-\ddot a/a\dot a\right)\int_0^tadt$
are both in the range $0< K_\bullet \le 1$ and reflect the reduction of growth
of perturbations caused by the cosmological constant. The scale factor of the
background model  is given by
$$ a(t)= \left({\Omega_{m} \over 1-\Omega_{m}} \right)^{1 \over 3}
\sinh^{2 \over 3}\left({3H_{0}t\sqrt{1-\Omega_{m}}} \over 2 \right)$$  
Here $H_0\equiv (\dot a/a)(t_0) =\dot a (t_0)$ is the Hubble constant today.
The $K_\bullet$-factors go to $1$ when $t\ll t_0$ or when 
$\Omega_m \rightarrow 1$ ($\Omega_{\Lambda} \rightarrow 0$). At decoupling 
$K_{\delta}=K_{V}=1$. An analytical approximation for $K_{\delta}(t)/\Omega_m$
with sufficient accuracy can be found in \cite{car92} and for 
$K_{V}(t_0)/\Omega_m$  in \cite{la}.

The power spectrum of density fluctuations is given by
\bea 
P(k,t)&\equiv& <\delta(t,k) \delta^*(t,k)> = A_sk^{n_s}T_m^2(k;t)a^2(t)
   K^2_{\delta}(t)/\Omega_m^2,\\ \nonumber
A_s &=& 2\pi^{2}\delta_{h}^{2}(3000{\rm Mpc}/h)^{3+n_s},
\label{pkz}
\eea
where $T_m(k,t)$ is transfer function (divided by the growth factor) and  
$\delta_h$ is the present matter density perturbation at horizon scale. We use 
the analytical approximation of   $T_m(k,t)$ in the space of cosmological 
parameters $h$, $\Omega_m$, $\Omega_b$, $\Omega_{\Lambda}$,
 $\Omega_{\nu}$ and $N_{\nu}$ (number of species of massive neutrino) 
by Eisenstein \& Hu (1999). 

From Eq.~(\ref{ani}), taking into account adiabaticity and setting 
$\bx_0={\bf 0}$, we obtain
\be
{\Delta T\over T}({\bf n})={1\over 3}\Phi(\eta_{dec},{\bf n}\eta_{0})+
2\int_{0}^{\omega_{e}}{\partial \Phi(\eta_0-\om,{\bf n}\om )
	\over\partial\eta}d\omega+
n_{\alpha} V^{\alpha}(\eta_{dec},{\bf n}\eta_{0})+{1\over 3}\delta_m(\eta_{dec},{\bf n}\eta_{0}), 
\label{dt}
\ee
where {\bf n} is the unit vector in direction of the incoming photon and we 
have used $\bx(\eta)={\bf n}(\eta_0-\eta)$, $\bx_{dec}\simeq{\bf n}\eta_0$. 
The variable $\omega$ is the affine parameter along the geodesic
which begins at the observer and ends in the emission point at the 
last scattering surface. The present value of conformal times, $\eta_0$
gives also the present particle horizon or the distance to the last-scattering 
surface.
The first term in (\ref{dt}) is the well known Sachs-Wolfe effect (SW), 
the second term is the integrated Sachs-Wolfe effect (ISW) which is important
only at late times, where $K_{\delta}(t)$ starts to deviate from $1$ and
${\partial \Phi\over \partial \eta} \ne 0$, 
the third is the Doppler term (D) and the last is the acoustic term (A).
At large angular scales ($\approx 10^{o})$, where anisotropies have been 
measured by COBE \cite{ben96}, the SW and ISW effects dominate. However,
if we want to calculate $C_{10}$ with good accuracy, we must to also
take into account the other terms.
The angular correlation function of $\Delta T/T$ can be written
symbolically as
\bea
\nonumber
<{\Delta T\over T}({\bf n_1})\cdot {\Delta T\over T}({\bf n_2})>={\rm <SW\cdot SW>+2<SW\cdot ISW>+<ISW\cdot ISW>}\\
{\rm +<A\cdot A>+2<SW\cdot A>+<D\cdot D>}~.
\label{dtcf}
\eea
The cross-correlators ${\rm <D\cdot SW>}$ and ${\rm <D\cdot A>}$ are omitted 
because they are strongly suppressed on large angular scales. 
Indeed, if one uses Fourier presentations  for the variables (\ref{sol}) in the
equations (\ref{dt}-\ref{dtcf}) one finds that the $k$-integrand of
these terms contains a spherical Bessel function $j_1(k\eta_0(\bn_1-\bn_2))$
which oscillates for large angular separations, strongly reducing the integral
if compared to the ${\rm <SW\cdot A>}$ term where the integrand has a 
definite sign.
The terms ${\rm <ISW\cdot A>}$ and ${\rm <ISW\cdot D>}$ are also omitted 
because the ISW effect gives the maximal contribution to $\Delta T/T$ at 
the largest angular scales of the range of interest (at lowest spherical 
harmonics) where A and D are nearly zero. 
At 'smaller' angular scales ($\ell\approx 10$) where contribution of A 
and D  are not negligible, the ISW effect is very small. Therefore, their 
cross-correlation terms are very small.

We develop the ${\bf n}$-dependence of  ${\Delta T\over T}({\bf n})$  
in spherical harmonics 
$$
{\Delta T\over T}({\bf n})=\sum_{\ell,m}a_{\ell m}(\eta_0)Y_{\ell m}({\bf n}),
 \;\;< a_{\ell m} a_{\ell^{\prime}m^{\prime}}^*>= \delta_{\ell m}
	\delta_{\ell^{\prime}m^{\prime}}C_\ell~.
$$
The CMB power spectrum, $C_\ell$, has the same components as the
correlation function:
\be
C_\ell=C_\ell^{\rm SW}+ C_\ell^{\rm SW-ISW}+ C_\ell^{\rm ISW}+ C_\ell^{\rm A}+ C_\ell^{\rm SW-A}+ C_\ell^{\rm D}
\label{cl}
\ee
Each component on the right hand side comes from the corresponding 
contribution to $\Delta T$ above and is proportional to $ \delta_h^2$. 
Using the solutions (\ref{sol}) we obtain analytic approximations for them.

We first approximate the SW and ISW contributions  in the form
\be
C_\ell^{\rm SW+ISW}\equiv C_\ell^{\rm SW}+ C_\ell^{\rm SW-ISW}+ 
C_\ell^{\rm ISW}=K^2_\ell C_\ell^{\rm SW},
\ee
where the factors $K_\ell$ ($\ge 1$) take into account the contribution 
of the ISW effect for each spherical harmonic. They have been calculated 
by Kofman \& Starobinsky (1985) and Apunevych \& Novosyadlyi (2000) for 
different $\Lambda$-models. Instead of
the direct time consuming calculations of the ISW contribution, we use
the following analytic approximations:
\bea
\nonumber
K^2_2=1+8.20423\times\exp(-\Omega_m/0.01157)+
3.75518\times\exp(-\Omega_m/0.13073),\\
\nonumber
K^2_3=1+2.25571\times\exp(-\Omega_m/0.03115)+
2.35403\times\exp(-\Omega_m/0.15805),\\
\nonumber
K^2_4=1+1.80309\times\exp(-\Omega_m/0.0323)+
1.88325\times\exp(-\Omega_m/0.16163)\\
\nonumber
{\rm and}\;\;\; K^2_\ell=1+[23.46523\times\exp(-\Omega_m/0.0122)+
11.03227\times\exp(-\Omega_m/0.14558)]/(\ell+0.5)
\eea
for $\ell\ge 5$. These approximation formulae are determined from
 the data presented in the tables of \cite{ks85} and \cite{an00}. 

Using solutions (\ref{sol}) and the definition of the density power 
spectrum~(\ref{pkz}) we obtain the following general expression for the SW 
contribution to the CMB power spectrum:
\be
C_\ell^{\rm SW}={\pi\eta_0^{n_s-1}\delta_h^2\over 2^{n_s-1}D^2(t_0)}
\int_0^\infty dk k^{n_s-2}T_m^2(t_{dec},k)j_\ell^2(k\eta_0),
\label{clswi}
\ee
where $T_m^2(t_{dec},k)$ is the transfer function of matter density 
perturbations at decoupling, $D(t_0)\equiv K_{\delta}(t_0)/\Omega_m$ is 
the value of the 
growth factor at the current epoch, and $j_\ell$ is the spherical Bessel 
function of order $\ell$. For reasonable values of spectral index 
$-3\le n_s \le 3$ the main contribution to the integral~(\ref{clswi}) comes 
from very small $k$ where $T_m(t_{dec},k)\approx 1$ and can be 
omitted. Then integral can be performed analytically and the result can be
expressed in terms of $\Gamma$-functions:
\be
C_\ell^{\rm SW}={\pi^2\delta_h^2\over 8D^2(t_0)}{\Gamma(3-n_s)
\Gamma(\ell+{n_s-1\over 2})\over \Gamma^2(2-n_s/2)\Gamma(\ell+{5-n_s\over 2})}.
\label{clsw}
\ee
In the same way we obtain the expressions for the other components of 
equation (\ref{cl}):
\bea
C_\ell^{\rm A}={\pi\eta_0^{n_s+3}\delta_h^2a^2(t_{dec})\over 18\cdot2^{n_s}D^2(t_0)\Omega_m^2}\int_0^\infty dk 
k^{n_s+2}T_b^2(t_{dec},k)j_\ell^2(k\eta_0),\\
\label{cla}
C_\ell^{\rm SW-A}=-{\pi\eta_0^{n_s+1}\delta_h^2a(t_{dec})\over 3\cdot2^{n_s-1}D^2(t_0)\Omega_m}\int_0^\infty dk 
k^{n_s}T_m(t_{dec},k)T_b(t_{dec},k)j_\ell^2(k\eta_0),\\
\label{clswa}
C_\ell^{\rm D}={\pi\eta_0^{n_s+1}\delta_h^2a(t_{dec})\over 2^{n_s-1}D^2(t_0)\Omega_m}\int_0^\infty dk 
k^{n_s}T_b^2(t_{dec},k)j_\ell^{\prime 2}(k\eta_0),
\label{cld},
\eea
where $T_b(t_{dec})$ is transfer function for density perturbations of 
baryons \cite{eh98} and $(^{\prime})$ is the derivative w.r.t the argument 
$x=k\eta_0$. The minus sign in the expression for $C_\ell^{\rm SW-A}$ reflects 
the anti-correlation of the gravitational potential and density fluctuations:
large positive density fluctuations generate deep negative potential wells. 
If we set $T_m=T_b=1$, the integrals (\ref{cla} --\ref{cld}) diverge for 
all $\ell$  for $n_s\ge 1$  because the main contribution to integrals of 
the D and A terms comes from small scales. Hence here the transfer functions 
must be kept and the integrals have to be calculated numerically. Fortunately, 
the integrands decay rapidly for large wave numbers and  99.9\%  of the 
contribution comes from the range $0.001\le k\eta_{dec}\le 0.1$, so that the 
integration is not very time consuming. 

\begin{figure}
\plottwo{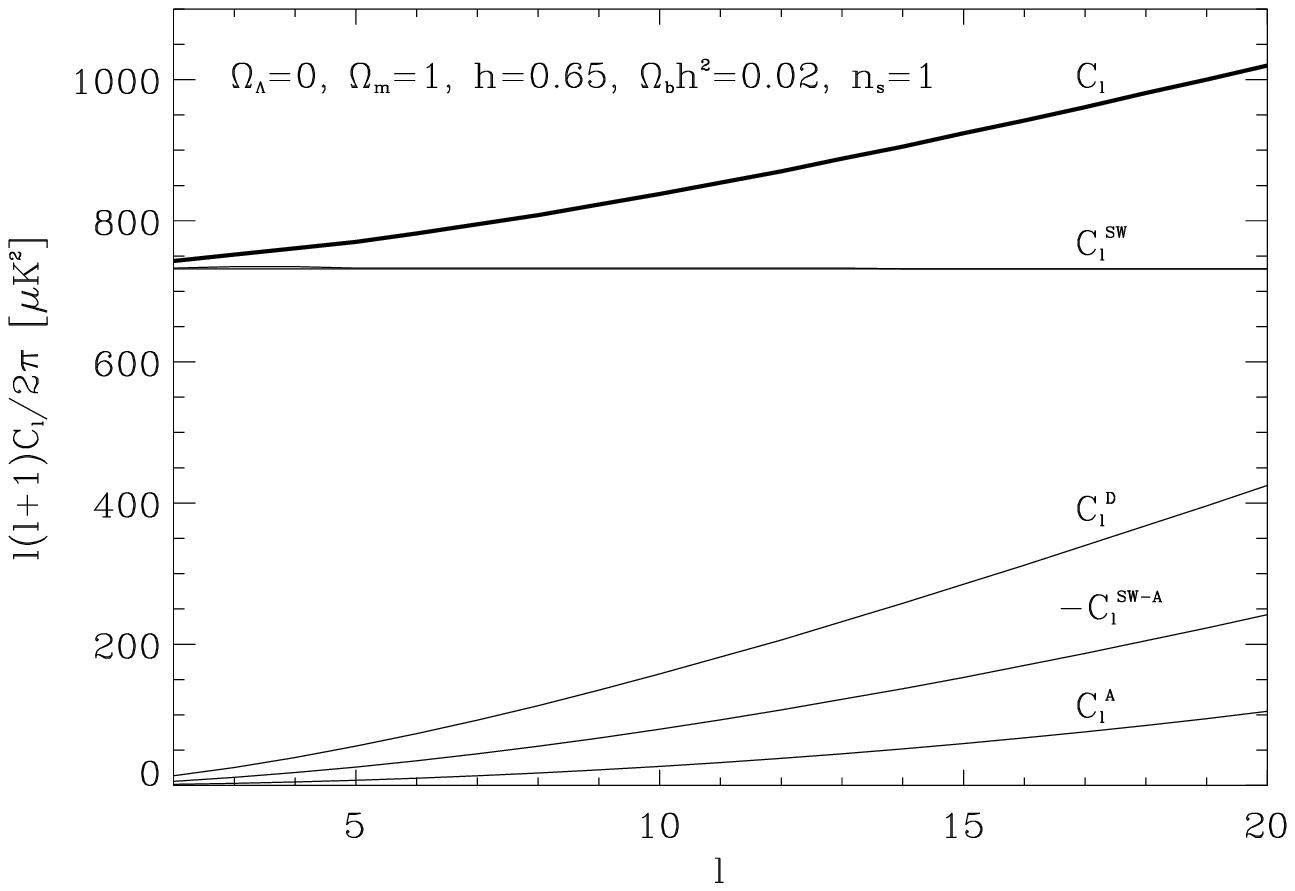}{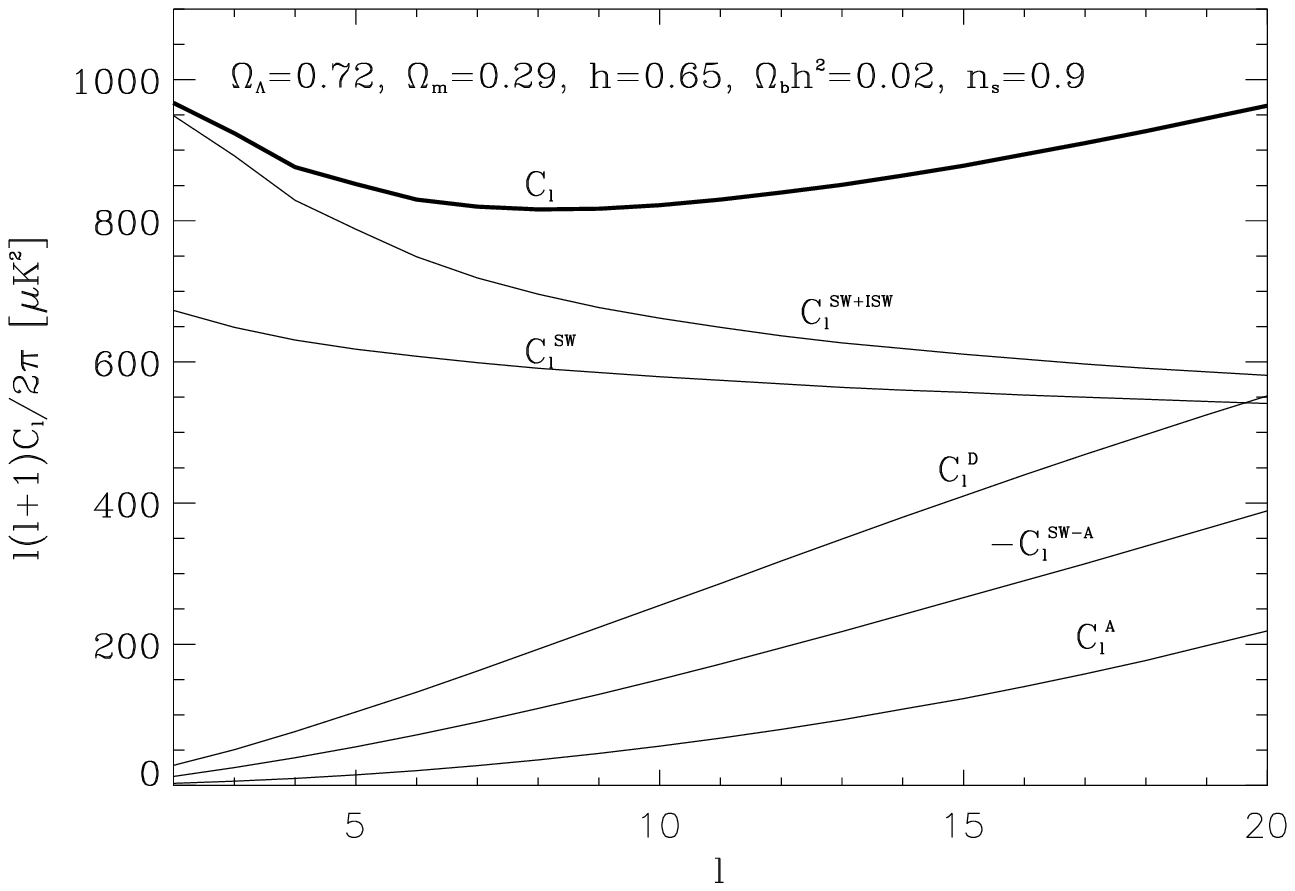}
\caption{The CMB power spectrum and the different contributions discussed 
in the text (formulae \ref{clsw}-\ref{cld}) for a pure matter model 
(left panel) and a $\Lambda$ dominated model  (right panel). }
\label{fig_eff}
\end{figure}

In Fig.\ref{fig_eff} the CMB power spectrum $C_\ell$ at large angular scales 
($\theta\ge 20^o,\;\;\ell\le 20$) together with the  contributions from 
the different terms given in (\ref{clsw}-\ref{cld}) is shown 
for a pure matter  and a $\Lambda-$ dominated model.
The relation of the contributions from different terms at $\ell=10$ are
$$
C_\ell^{\rm SW}:C_\ell^{\rm A}:C_\ell^{\rm SW-A}:C_\ell^{\rm D}=1:0.04:-0.11:0.22
$$
for the matter dominated flat model ($\Omega_m=1$) and 
$$
C_\ell^{\rm SW}:C_\ell^{\rm A}:C_\ell^{\rm SW-A}:C_\ell^{\rm D}=1:0.08:-0.23:0.39.
$$
for the $\Lambda$ dominated model with the cosmological parameters shown in 
the figure. Therefore, a few percent accuracy of the normalization to 
4-year COBE $C_{10}$ data can be achieved only if all these effects 
are taken into account. 

\begin{figure}
\plottwo{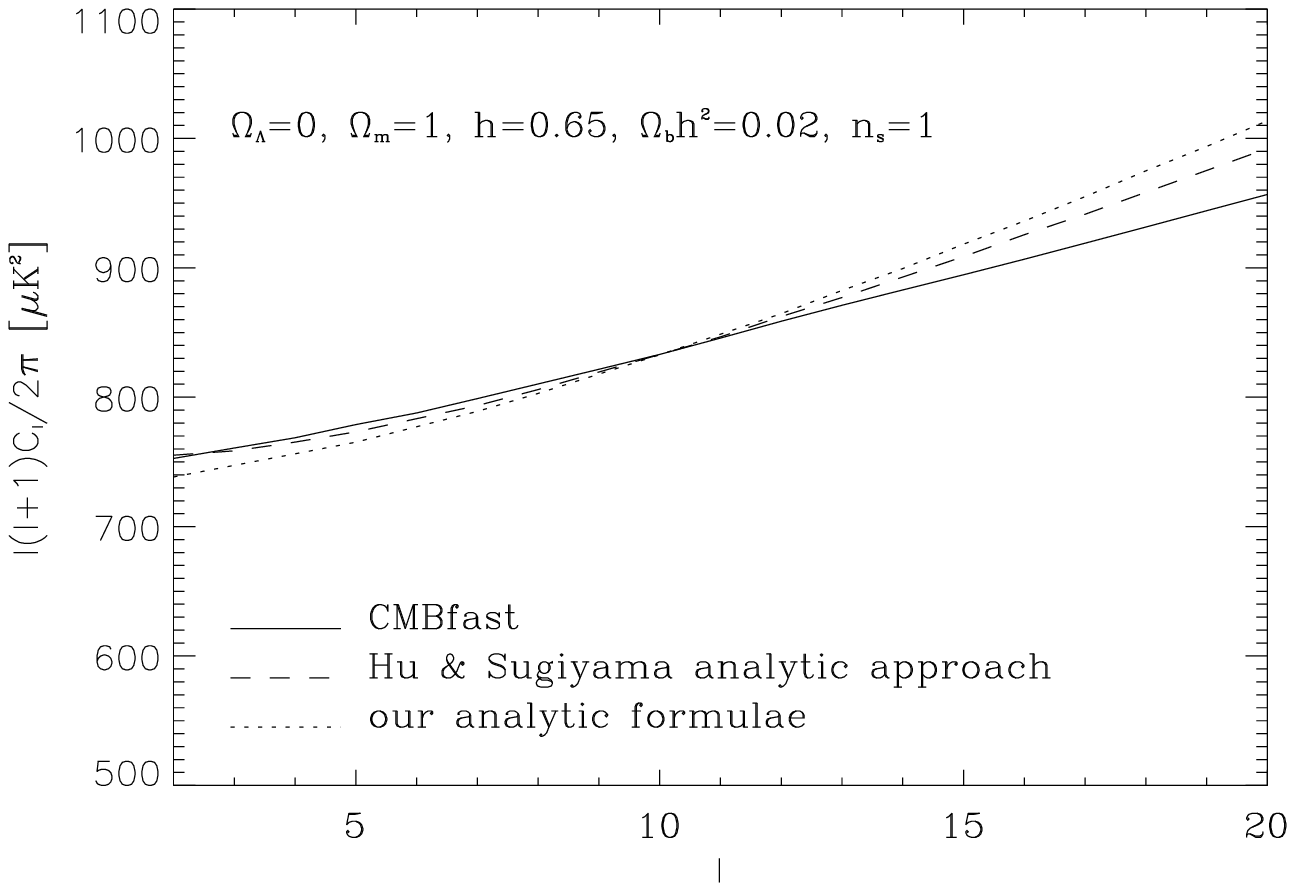}{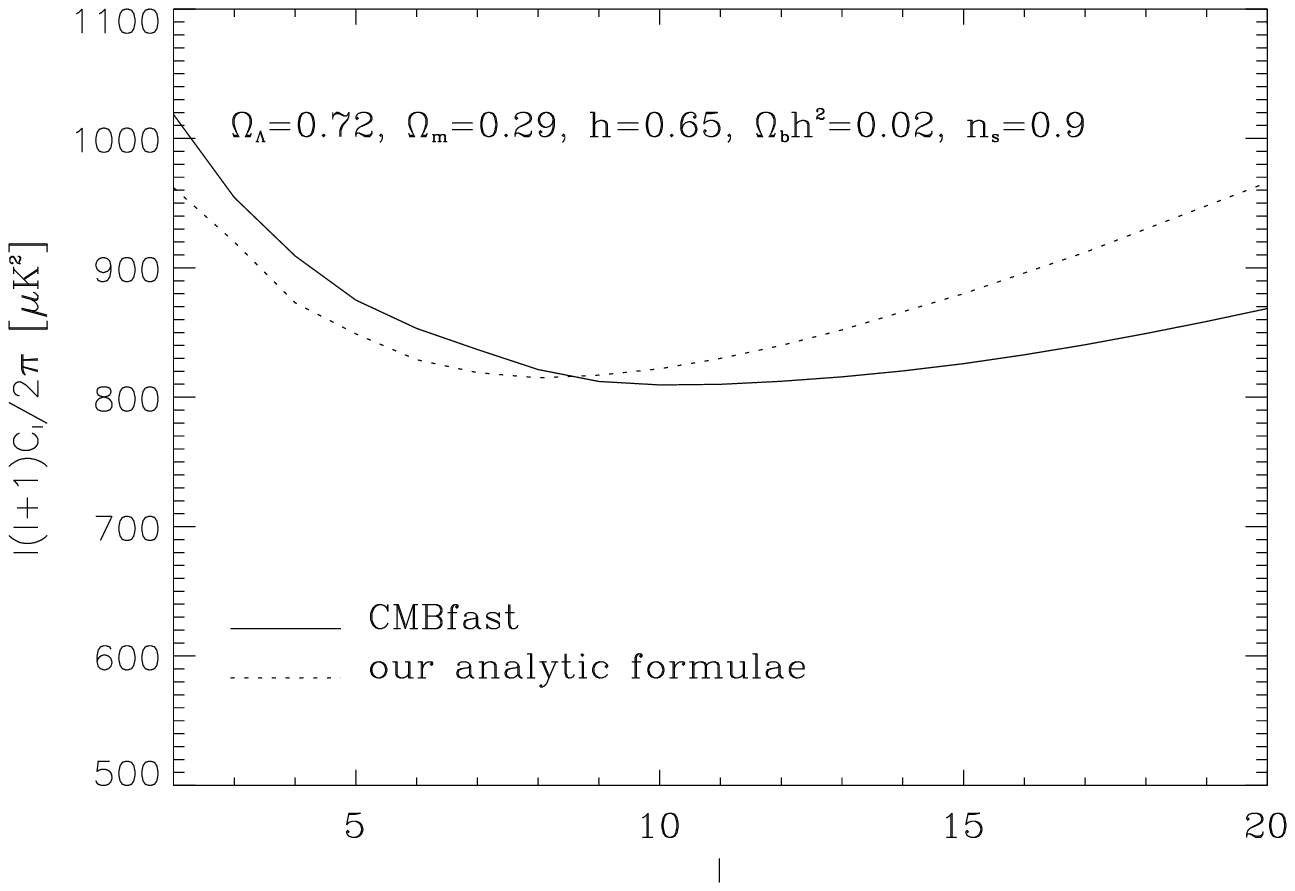}
\caption{The CMB power spectrum at COBE scales calculated by CMBfast
(solid line) and by our analytical formulae (\ref{clsw}-\ref{cld}) 
(dotted line). For the pure matter model we also show the spectrum 
calculated with the analytic approach of \cite{hs95} (dashed line) (this 
approach does not allow a cosmological constant). All spectra are normalized 
to the best fit for  $C_{10}$ from the 4-year COBE data given in \cite{bw97}.}
\label{fig_cl}
\end{figure}

In Fig. \ref{fig_cl}  the CMB  power spectrum  at large scales
calculated using the analytic formulae (\ref{clsw}-\ref{cld}) and using
CMBfast  are shown for comparison. 
In the left panel we also present the power spectrum calculated by 
the analytical approach of \cite{hs95} (renormalized to the CMBfast value 
of $C_{10}$). 

The calculations show that value of $C_{10}$ calculated by our method 
deviates from the value calculated with CMBfast by 0.5\% for the matter 
dominated flat model ($\Omega_m=1$) and 2.7\% for the $\Lambda$ dominated 
model ($\Omega_m=0.2$). Therefore, our analytic approach is sufficient to 
normalize fast  the power spectrum of scalar perturbations 
to the 4-year COBE data with virtually the same precision as CMBfast, 
the difference is less than $3\%$. (Remember, that the experimental
errors of the  COBE data are about $ 14\%$, so that 
the best-fit normalization parameter $C^{COBE}_{10}$ has the same error.)

Another comparison of our normalization procedure with CMBfast comes from
the value of $\sigma_8$. For the flat model (left panel of 
Fig.~\ref{fig_cl}) our approximation for the normalization together with
the analytical transfer function of Eisenstein \& Hu (1998; 1999) leads to 
$\si_8=1.58$, the corresponding value calculated from CMBfast is 1.53.
For the $\Lambda$ dark matter model (right panel) our $\sigma_8=0.62$, while  
CMBfast gives $\sigma_8=0.64$. The agreement of both approaches is
quite well (the 5\% difference includes also the errors in the  approximation
of the transfer function  which is actually of this order).

The slight deviation of $C_{10}$ as calculated by our code from the value 
obtained with CMBfast ($\le 3\%$) in spite of using the same analytic 
best-fit formula for $C_{10}^{COBE}$ by \cite{bw97} is due to a difference 
in the form of the  spectra as shown in the Fig.~\ref{fig_cl}. This difference 
grows when $\Omega_m$ decreases. There are several possible reasons for this
deviation in the form of the CMB power spectrum in our analytic approach 
from the exact numerical calculation:
 1) We have used the solutions for the 
evolution of density, velocity and gravitational potential perturbations in 
the $\Lambda$- dust Universe. In reality, at decoupling the role 
of radiation is not completely negligible, this slightly influences the 
dynamics of the scale factor and the evolution of perturbations.
It also results in additional time dependence of gravitational 
potential (early integrated Sachs-Wolfe effect) which is not taken into 
account here.
2) Our approach does not take into account the effects of the collisionless 
dynamics of photons and neutrinos after decoupling. Especially, the induced
anisotropic stresses lead to $\approx 10\%$ difference of the gravitational 
potentials in the radiation-dominated epoch which results into a corrections 
of a few percent in the $C_\ell$'s. 
3) Instantaneous recombination and tight coupling which were assumed, also 
cause slight inaccuracies. They
should, however, be extremely small on the angular scales considered here. 
4) To calculate the terms $C_\ell^{A}$, $C_\ell^{SW-A}$ and $C_\ell^{D}$ 
we have used the analytic approximations for the transfer functions,
$T_m(t_{dec},k)$ and $T_b(t_{dec},k)$,
by  Eisenstein \& Hu (1998; 1999) which have an  accuracy $\sim 5\%$.

More details on the theory of CMB anisotropies can be found in the 
reviews by Durrer \& Straumann (1999) and Durrer (2001). 

\newpage

\section{Analytic formulae for  the amplitudes  and  locations
of acoustic peaks and dips in the CMB power spectrum}

For completeness, we repeat here the formulas used in our parameter search 
which can also be found in the cited literature.
 
We assume the standard recombination history and define the redshift of decoupling
$z_{dec}$ as the redshift at which the optical depth of Thompson scattering is unity.
A useful fitting formula for $z_{dec}$ is given by \cite{hs96}:
\be
z_{dec}=1048[1+0.00124\omega_b^{-0.738}][1+g_1\omega_m^{g_2}],
\label{zdec}
\ee
where
$$
g_1=0.0783\omega_b^{-0.238}[1+39.5\omega_b^{0.763}]^{-1},\;\;\;
g_2=0.56[1+21.1\omega_b^{1.81}]^{-1},
$$
$\omega_b\equiv\Omega_bh^2$ and $\omega_m\equiv\Omega_mh^2$.

\subsection{Locations}

The locations of the acoustic peaks in the CMB power spectrum depend on 
the value of sound horizon  at decoupling  
epoch $r_s(\eta_{dec})\equiv \int_{0}^{\eta_{dec}}d\eta^{\prime}c_s$  
and the angular diameter distance to the last scattering surface,
 $d_A(z_{dec})$. Comparing with numerical calculations it was shown 
(see \cite{efs99,hu01,dl01} and references therein) 
that the  spherical harmonic which corresponds to the $m$-th acoustic 
peak is well approximated by the relation
\bea
\ell_{p_m}= (m-\phi_m)\pi{d_A(z_{dec})\over r_s(z_{dec})},
\label{lpm}
\eea
where $\phi_m$ take into account the shift of $m$-th peak
from its location in the idealized model which is
caused by driving effects from the decay of the gravitational potential. 
Doran  and Lilley (2001) give an accurate analytic approximation in the form
\be
\phi_m={\bar\phi} - \delta\phi_m~,\ee
where ${\bar\phi}$ is overall phase shift of the spectrum (or the first 
peak) and $\delta\phi_m$ is a relative shift
of each peak and dip caused by the Doppler shift of the oscillating fluid.
For the overall phase shift of the spectrum they find
\be
{\bar\phi}=(1.466-0.466n_s)a_1r_*^{a_2},
\ee  
where 
$$r_*\equiv \rho_{rad}(z_{dec})/\rho_{m}(z_{dec})={0.0416\over \omega_m}
\left({1+\rho_{\nu}/\rho_{\gamma}\over 1.6813}\right)
\left({T_0\over 2.726}\right)^4\left({z_{dec}\over 1000}\right)$$
is the ratio of radiation to matter  at decoupling, and 
$$
a_1=0.286+0.626\omega_b\;,\;\;a_2=0.1786-6.308\omega_b+174.9\omega_b^2-1168\omega_b^3
$$
are fitting coefficients. Here and below the numbers in the expressions are 
obtained for a present CMB temperature of $T_0=2.726$K and the ratio of 
densities of massless neutrinos and photons 
$\rho_{\nu}/\rho_{\gamma}=0.6813$ for three  massless neutrino species 
(correspondingly $f_{\nu}\equiv \rho_{\nu}/(\rho_{\gamma}+\rho_{\nu})=0.405$).
All values can be easily scaled to other values of $T_0$ and $f_{\nu}$.

The relative shift of the 1st acoustic peak is zero, $\delta\phi_1=0$.
For the 2nd one it is 
\be
\delta\phi_2=c_0-c_1r_*-c_2/r_*^{c_3}+0.05(n_s-1)~,
\ee
with
$$c_0=-0.1+0.213e^{-52\omega_b},\;\;\;c_1=0.015+0.063e^{-3500\omega_b^2},\;\;
c_2=6\cdot 10^{-6}+0.137(\omega_b-0.07)^2,\;\;\;c_3=0.8+70\omega_b,$$
and for the 3rd peak
\be
\delta\phi_3=10-d_1r_*^{d_2}+0.08(n_s-1)~,
\ee
with
$$d_1=9.97+3.3\omega_b,\;\;\;d_2=0.0016+0.196\omega_b+2.25\cdot 10^{-5}\omega_b^{-1}.$$

The formula (\ref{lpm}) is correct also for the location of dips if we set 
 $m=3/2$  for the 1st dip and $m=5/2$ for the 2nd dip.
The relative shift of the first dip given by \cite{dl01} is
\be
\delta\phi_{3/2}=b_0+b_1r_*^{1/3}\exp{b_2r_*}+0.158(n_s-1)~
\ee
with 
$$b_0=-0.086-2.22\omega_b-140\omega_b^2~,\;\;b_1=0.39-18.1\omega_b+440\omega_b^2,\;\;
b_2=-0.57-3.8\exp({-2365\omega_b^2})~.$$

The angular diameter distance to the last scattering surface is given by
\be
d_A(z_{dec})={c\over H_0\sqrt{|\Omega_k|}}\chi(\eta_0-\eta_{dec})~,
\label{da}
\ee
where $\chi(x)=x,\;\;\sin{x}\;\;{\rm or}\;\;\sinh{x}$ for flat, closed or 
open models respectively, and
\be
\eta_0-\eta_{dec}=\sqrt{|\Omega_k|}\int_0^{z_{dec}}{dz\over \sqrt{\Omega_{rad}(z+1)^4+
\Omega_m(z+1)^3+\Omega_{\Lambda}+\Omega_k(z+1)^2}}~.
\ee
Since, the sound speed in the pre-recombination plasma is 
\be
c_s=c/\sqrt{3(1+R)}\;\;\;{\rm with}\;\;\;  
R\equiv 3\rho_b/4\rho_{\gamma}=30315(T_0/2.726)^{-4}\omega_ba
\ee
 and scale factor is well approximated by
\be
a(\eta)=a_{eq}\left({\eta\over \eta_1}+({\eta\over 2\eta_1})^2\right),
\ee
with
$$a_{eq}={4.16\cdot 10^{-5}\over \omega_m}\left({1+\rho_{\nu}/\rho_{\gamma}\over 1.6813}\right)
\left({T_0\over 2.726}\right)^4,\;\;\;
\eta_1\equiv {\eta_{eq}\over 2(\sqrt{2}-1)},$$
the integral for sound horizon can be reduced to the analytic formula
\be
r_s(\eta_{dec})={19.9\over \sqrt{\omega_b\omega_m}}\left({T_0\over 2.726}\right)^2
\ln{\sqrt{1+R_{dec}}+\sqrt{R_{dec}+R_{eq}}
\over 1+\sqrt{R_{eq}}}\;\;\rm Mpc.
\label{rs}
\ee

The deviation of the acoustic extrema locations calculated using formulae
(\ref{lpm}-\ref{rs}) from the values obtained by CMBfast code is 
$< 3\%$ for a sufficiently wide range of parameters.

\subsection{Amplitudes}

The amplitude of the 1st acoustic peak can be approximated by the following 
expression
\be
A_{p_1}= {\ell_{p_1}(\ell_{p_1}+1)\over 2\pi} \left[C^{SW}_{\ell_{p_1}}
+C^{SW}_2 {\tilde{\cal A}}(\Omega_b,\Omega_{cdm},\Omega_k,n_s,h) 
\right]~,  
\label{ap1}
\ee
where 
\be
{\tilde {\cal A}}\equiv 0.838{\cal A}=
\exp{[{\tilde a_1}+a_2\omega_{cdm}^2+
a_3\omega_{cdm}+a_4\omega_b^2+a_5\omega_b+
a_6\omega_b\omega_{cdm}+a_7\omega_k + a_8\omega^2_k+a_9(n_s-1)]}
\label{A}
\ee 
and $C^{SW}_{l_{p_1}}$ is given by (A9).
We have re-determined the best-fit coefficients $a_{i}$ using the 
values of the 1st acoustic peak amplitudes from CMBfast for the grid of 
parameters given below. Their values are
\bea
\nonumber
&{\tilde a_1}=2.326,\;\; a_2=8.906,\;\;a_3=-7.733,\;\;a_4=-115.6,\;\;a_5=35.66,&\\
&a_6=-7.225,\;\;a_7=1.96,\;\;a_8=-11.16,\;\;a_9=4.439.& 
\label{ai}
\eea
The deviations of this approximation  from the numerical value obtained 
by CMBfast  are $\le 5\%$ within the range of cosmic parameters,
$0.2\le\Omega_m\le 1.2$, 
$0\le\Omega_{\Lambda}\le 0.8$, $0.015\le\Omega_b\le 0.12$, 
$0.8\le n_s\le 1.2$ and $0.4\le h\le 1.0$. 
 
To calculate the  amplitudes of the 2nd and 3rd peaks we use 
the relations $$H_2\equiv \left[\ell_{p_2}(\ell_{p_2}+1)C_{\ell_{p_2}}\right]/
\left[\ell_{p_1}(\ell_{p_1}+1)C_{\ell_{p_1}}\right]\;\;\;{\rm and}\;\;\; 
H_3\equiv \left[\ell_{p_3}(\ell_{p_3}+1)C_{\ell_{p_3}}\right]/
\left[\ell_{p_1}(\ell_{p_1}+1)C_{\ell_{p_1}}\right]$$
given by Hu et al. (2001). This leads to the following amplitudes
\bea
A_{p_2}=A_{p_1}H_2(\Omega_m,\Omega_b,n_s),\;\;\;
{\rm with}\;\;\; H_2={0.925\omega_m^{0.18}2.4^{n_s-1}\over \left
[1+\left(\omega_b/0.0164\right)^{12\omega_m^{0.52}}\right]^{1/5}}~,\\
A_{p_3}=A_{p_1}H_3(\Omega_m,\Omega_b,n_s),\;\;\;
{\rm with}\;\;\;H_3={2.17\omega_m^{0.59}3.6^{n_s-1}\over  
\left[1+\left(\omega_b/0.044\right)^2\right]\left[1+1.63(1-
 \omega_b/0.071)\omega_m\right]}.
\eea

This approximation for $A_{p_3}$ deviates  by less than 5\% from the value 
obtained with CMBfast for parameters within the range specified above. The 
accuracy of $A_{p_2}$ is better than 9\%. For some parameter values the 
second peak is under estimated leading to this somewhat poorer accuracy.

\end{document}